\documentclass{aa} 

\usepackage{natbib}
\usepackage{graphicx}
\usepackage{txfonts}
\usepackage{hyperref}

\usepackage{amstext}
 
\begin{document} 

  \title{Revised 
  orbital parameters of the $\gamma^2$ Velorum system\thanks{Based 
  on observations collected at the European Southern Observatory under ESO program 112.25Z3.001}
        }
  
\titlerunning{Orbital parameters of $\gamma^2$ Velorum}
\author{Werner Schmutz\inst{1} \corrauth{werner.schmutz@pmodwrc.ch}\and
Christian A. Hummel\inst{2} \and 
Gloria Koenigsberger\inst{3} \and 
Florentin Millour\inst{4} \and 
Joel Sanchez-Bermudez\inst{5} 
          }

  \institute{Physikalisch-Meteorologisches Observatorium Davos and World Radiation Center,
              Dorfstrasse 33, CH-7260 Davos Dorf, Switzerland
              \and 
            European Southern Observatory, Karl-Schwarzschild-Stra\ss e 2, 85748 Garching, Germany
         \and
            Instituto de Ciencias F\'isicas, Universidad Nacional Aut\'onoma de M\'exico, Ave. Universidad S/N, Cuernavaca, 62210 Morelos, M\'exico
         \and
            Observatoire de la C\^ote d'Azur, France
         \and
            Universidad Nacional Aut\'onoma de M\'exico. Instituto de Astronom\'ia. A.P. 70-264, Ciudad de M\'exico, 04510, M\'exico
             }

  \date{Received March 24, 2026; accepted June 23, 2026}

  \abstract
   {$\gamma^2$ Velorum is the closest and visually brightest Wolf-Rayet binary system. Its eccentric orbit modulates the X-rays observed from the wind-wind interaction, and its 
   large separation
   allows for spatially resolving both components. }
   {We aim to  strengthen the constraints on $\gamma^2$ Velorum's properties and, in particular, solve the discrepancy between the eccentricity determined from the emission lines and that from the absorption lines.}
   {We obtained VLT/GRAVITY observations and combined them with earlier spatially-resolved data at different orbital phases.}
   {Strong constraints on all orbital parameters were determined and, in 
   particular, we find that $e=0.322$, close to what was derived from the emission lines.
   The X-ray light curve declines as $s^{-3}$ after periastron, where $s$ is the separation of the two stars, but its modulation is likely affected by absorption and occultation of the X-ray emitting region at other orbital phases.  We find that previous discrepancies in the reddening value can be traced to a brighter K-band magnitude than that predicted by the WR wind models. We conclude $E(B-V)=0.02\pm 0.02$\,mag.  Our now more precise mass and radius values combined with previously determined effective temperatures  provide  very strong constraints on evolutionary models. The closest match for the O-star is provided by an initial mass $M_\mathrm{init}=28.7\,\mathrm{M}_\odot$ rotationally mixed model and a $M_\mathrm{init}=32\,\mathrm{M}_\odot$ model for the WR star, with negligible accretion onto the O-star during the WR progenitor's Roche Lobe overflow phase. 
   However, the temperature of the O star is higher and the mass of the WR star is found to be smaller than predicted by the evolutionary tracks for the current epoch, consistent with the well-known  "mass-discrepancy problem" in massive stars.
}  
{$\gamma^2$ Velorum's strongly constrained parameters now make it ideal  for testing structure, evolution and stellar wind models. }

  \keywords{stars: binaries: eclipsing-stars -- stars: individual: $\gamma^2$ Velorum; HD\,68273;  -- stars: binaries: close }

\maketitle
\nolinenumbers 

\section{Introduction}
The binary $\gamma^2$ Velorum (HD\,68273) contains the closest and visually brightest Wolf-Rayet star (WR) and it is the closest example of a binary system in which the winds of two massive stars collide. As such, it offers a unique opportunity for firmly establishing the stellar and orbital properties, which, in turn,  provide strong constraints on evolutionary models. In addition, thanks to the eccentricity of the orbit, the varying separation between the two stars allows examination of the manner in which wind-wind collision processes depend on orbital separation.
The separation between two stars depends on the eccentricity of the orbit,  a parameter that in WR+O binaries is challenging to determine very accurately for several reasons.  First,  the WR wind regions from which emission lines arise are gravitationally detached from the star, connected to it only through the radially outward-acting radiation pressure (see Fig.,11 of \citet{Schmutz1997}), thus it is not clear whether the radial velocities that are derived from these lines actually reflect the orbital motion.  Second, the presence of a WCR introduces an asymmetry in the WR wind, which again can distort the measured radial velocity curve.  Finally, although the absorption lines arising in the O-star companion are better suited to trace the orbit, the much stronger and variable emissions arising in the system can distort their profiles, leading to uncertainties in the radial velocity measurements that are not straightforward to quantify.

$\gamma^2$\,Vel is a system that highlights the above difficulties.
\citet{Schmutz_etal1997} obtained  $\gamma^2$ Vel's radial velocity orbit from high-resolution spectra covering almost two orbital periods ($P=78.5$ d) with a one-day time resolution. Although the average radial velocity curve over all the emission lines in \citet{Schmutz_etal1997} yielded $e=0.33$, 
the absorption lines led to a bimodal result in  which several absorption lines gave $e=0.28$ while others were grouped around $e=0.38$.  
Richardson et al. (2017) presented orbital elements for 10 emission lines with eccentricities ranging from 0.118 to 0.423.  In both studies, it is suggested that the wind collision region is responsible for distorting to different degrees the radial velocity curves of different spectral features. Radial velocity curves are obtained by measuring the centroid of spectral lines over the orbital cycle.  For such a RV curve to represent the orbital motion, the line shape must remain constant.  However, the WCR produces a cavity in the Wolf-Rayet wind which  causes the line shape to undergo phase-dependent variations.  The effect, primarily for UV P Cygni lines, was modeled by \citet{2004A&A...423..267G}.   In addition, the WCR can add line flux from the outflowing shocked material.  The presence of narrow sub-peaks superposed on the broad WR lines is often interpreted as arising in the WCR with models of the expected sub-peak velocity as a function of orbital phase constructed based on simplifying approximations \citep{Luehrs_1997, 2000MNRAS.318..402H, 2009MNRAS.395..962I, 2021MNRAS.503..643W}.  However, the manner in which these sub-peaks affect the overall RV curve is difficult to assess because the flux of optical line emission from the WCR, though likely small compared to that of the WR wind, is not known.  

Richardson et al. (2017) show results from numerical simulations of $\gamma^2$ Vel's  WCR structure and suggest that the observed excess emission and variable P Cyg absorption in several of $\gamma$ Vel's lines are compatible with the simulation results. Unfortunately, these authors did not perform the radiative transfer calculation needed to compute the line profiles from which synthetic RV curves could be derived.  Hence, although it is suspected that the wide range of orbital parameters obtained from different spectral lines in WR binaries is caused by colliding winds, it is not clear which lines produce RV curves that are closest to representing the actual orbital motion. 

$\gamma^2$ Vel  is located at a distance of only 336\,pc, allowing for interferometric observations that can eliminate this ambiguity in the eccentricity derived from radial velocity curves.  
In fact, $\gamma^2$ Vel  was the first binary ever resolved with interferometry \citep{1970MNRAS.148..103H} and its  importance can not be over stated. Not only is it the closest WR binary, it is also the shortest-period system that can be resolved with current observational infrastructure.  
The other WR binaries with resolved orbits are WR 137, WR138, WR140, all having orbital periods in the range 4-13 years \citep{Richardson_etal2016, Richardson_etal2024, 2021MNRAS.503..643W, Holdsworth_etal2024}.
The WWC conditions in such wide binaries differ significantly from those in the shorter-period  systems, where radial velocity curves are the only method for determining their orbits.   One more system that has been interferometrically resolved is WR133 \citep{2021ApJ...908L...3R} which has a 113\,d orbital period, but it is a nitrogen-rich WR.  With its 79\,d period, $\gamma$ Vel is thus the best system currently available to bridge the gap between the low-density colliding wind regime of the wide binaries and that of the shorter-orbit systems.

\citet{Lamberts_etal2017} measured the astrometric orbit from AMBER/VLTI measurements at 10 epochs. However, as their  astrometric measurements did not sufficiently constrain the eccentricity of the orbit, they adopted $e=0.326$ from the \citet{Schmutz_etal1997} velocity orbital solution, as well as $e=0.344$ from \citet{North_etal2007} interferometric measurements. Because the orientation of the binary orbit in the sky 
is such that the minor axis of the orbit is close to the line of nodes, and the three existing high-precision AMBER/VLTI measurements at phases 0.85-0.9 accurately  constrain this part of the astrometric orbital ellipse, the derived values of the orbital parameters $T_0$, $\omega_\mathrm{WR}$, $\Omega$, $i$, and the semi-major axis $a$ are only moderately affected by the uncertainty in the eccentricity. However, the impact is significant for anything that depends on the orbital eccentricity.  Specifically, the eccentricity value enters into the calculation of the orbital separation, which is a fundamental parameter for the calculation of the 
processes involved in the wind-wind interaction zone and in the resulting 
orbital phase-dependent X-ray luminosity.

The X-ray luminosity of a wind collision zone (WCZ) is predicted to depend on the inverse of the orbital separation if the shock is adiabatic \citep{Stevens_etal1992}. The current uncertainty in eccentricity maps into a range in the ratio of periastron to apastron distances $\rho = (1-e)/(1+e)$=0.56 to 0.40. Thus, despite $\gamma^2$ Vel's nearby location and relatively well-known parameters, the impact of the uncertainty in the eccentricity is such that the power of the X-ray luminosity dependence on separation has an uncertainty of the order of 25\,\%.

In this paper we report the results of new interferometric observations which solve the ambiguity in the eccentricity determination.  Section 2 describes the interferometric measurements.  In Section 3 we determine the astrometric orbit.  In Section 4 we refine the orbital period.  The distance to $\gamma^2$\,Vel is reviewed in Section 5.  
In Section 6 we discuss the X-ray variability in the context of the new eccentricity determination, 
investigate published radial velocity measurements,
reevaluate the stellar parameters, and compare its properties with evolutionary models. 
Section 7 summarizes our conclusions.
\section{Interferometric Measurements \label{Sec:2}}

\begin{table*}
\caption[]{Interferometric measurements\tablefootmark{a}
of $\gamma^2$\,Vel in the $K$\,band.
}
\label{tab:GRAVITY_observations}
\centering
\begin{tabular}{lcccccccccc} %
\hline
\noalign{\smallskip}
     & MJD     & phase & telescope     & $\rho$ & $\theta$ & $\Delta\alpha$ & $\Delta\delta$ & $\sigma_\mathrm{major\ axis}$  & $\sigma_\mathrm{minor\ axis}$ & $\theta_{\sigma_\mathrm{major\ axis}}$\\
Date & mid-obs &  & configuration & mas    & deg      &  mas           & mas            &  mas & mas & deg\\
\hline
\noalign{\smallskip}
13. Nov 2023 & 60261.316 & 0.139 & A0-G2-J2-J3 & 2.76 & 235.13    & $-2.26$        & $-1.58$        & 0.069 & 0.042 & 145.5\\
20. Nov 2023 & 60268.313 & 0.228 & A0-B5-J2-J6 & 3.65 & 229.25    & $-3.41$        & $-1.29$        & 0.074 & 0.034 & \ \ 71.9\\
14. Feb 2024 & 60354.048 & 0.319 & A0-G1-J2-K0 & 3.75 & 260.70    & $-3.70$        & $-0.61$        & 0.069 & 0.047 & 135.0\\
06. Mar 2024 & 60375.130 & 0.588 & A0-G1-J2-K0 & 1.96 & 325.72    & $-1.10$        & $\ \ 1.62$     & 0.067 & 0.042 & 179.5\\
\hline
\end{tabular}   
\tablefoot{
\tablefoottext{a}{MJD is the modified Julian date, the phase is calculated with Eq.\,1, $\rho$ is the separation between the stars, $\theta$ is the position angle of the WR star, $\Delta$x is the separation in right ascension, $\Delta$y is the separation in declination, and $\sigma_\mathrm{major\ axis}$, $\sigma_\mathrm{minor\ axis}$, and $\theta_{\sigma_\mathrm{major\ axis}}$ describe the 2.5-$\sigma$ uncertainty ellipse of the observation.
} 
}
\end{table*}

\begin{table}
\caption[]{Interferometric measurements\tablefootmark{a}
of $\gamma^2$\,Vel in the $K$\,band.
}
\label{tab:disk}
\centering
\begin{tabular}{lccc} 
\hline
\noalign{\smallskip}
     & $\Theta_\mathrm{UD}$(O) &  $\Theta_\mathrm{UD}$(WR) & $K_\mathrm{WR}-K_\mathrm{O}$\\
Date & mas                          &  mas             & mag \\
\hline
\noalign{\smallskip}
13. Nov 2023 & 0.96 & 0.87 & -0.43 \\
20. Nov 2023 & 1.06 & 0.72 & -0.24 \\
14. Feb 2023 & 1.06 & 0.88 & -0.35 \\ 
06. Mar 2023 & 1.37 & 0.99 & -0.40 \\
\hline
\end{tabular}   
\tablefoot{
\tablefoottext{a}{$\Theta_\mathrm{UD}$ are the diameters of the stars represented by a uniform disk and $\cal F_K$ are the stellar $K$-band fluxes. }
}
\end{table}

$\gamma^2$ Vel was observed in service mode with ESO's GRAVITY instrument, combining the light of the four auxiliary telescopes of the VLT Interferometer. The telescope configurations used were three times the large configuration, with baselines between the telescopes of 49\,m to 129\,m, and once the extended configuration, with a largest baseline of 202\,m. Before or after the 34\,min science observation, the visibility calibration target HD\,68512 (diameter of 1.30 milli-arcseconds, JSDC Catalogue) was observed for 28\,min. The data were reduced and calibrated with the GRAVITY pipeline to produce visibilities and closure phases as a function of wavelength. 

The visibilities basically look like those presented by \citet{Lamberts_etal2017}. There are peaks and troughs at wavelengths of WR emission lines, with visibility-peaks at the locations of C\,{\sc iv} emissions and troughs at wavelength locations of He\,{\sc i} and C\,{\sc iii} emissions.  
Therefore, for the interpretation of the interferometric data with a geometric model we masked out the WR emission lines by selecting continuum wavelength regions as shown in Fig.\,4 of \citet{Lamberts_etal2017} and fitted the continuum selection with a geometrical model of two uniform disks to calculate the relative positions of the two stars, the disk diameters and the brightness ratio in the K-band as reported in Tab.\,\ref{tab:GRAVITY_observations} and \ref{tab:disk}. We obtain apparent diameters that are comparable to those determined by \citet{Lamberts_etal2017}.

A simultaneous fit to all GRAVITY/VLTI data yields the stellar angular diameters
$\Theta_\mathrm{UD}$(O)$=1.08\pm 0.15$\,mas and 
$\Theta_\mathrm{UD}$(WR)$=0.87\pm 0.10$\,mas. 
It is not possible to determine whether the stellar diameters are different  for the four observing dates or whether the variations just reflect the uncertainties of the measurements. Therefore, the uncertainties given are the standard deviations of the four values. The 50\,\%\ larger standard deviation for the O star diameter might indicate that the measurements of the WR star are measurement uncertainties and those of the O star are apparent diameters that are different for the four observing dates. We note that according to the stellar diameter determined from the photometry in Sect.\,\ref{stellar_parameters}, $\Theta_\mathrm{flux}$(O)$=0.38$\,mas and $\Theta_\mathrm{flux}$(WR cont.)$=0.12$\,mas. Thus, the apparent diameters measured by interferometry are much larger. 

The larger the visibilities are, the smaller the emitting region at that wavelength. Thus, 
the observation that the visibilities have a peaked maximum at the wavelength of C\,{\sc iv} at 2.07\,$\mu$m, which is the strongest emission line in the K-band,  indicates that the continuum emitting regions of both stars are not resolved, because  continuum emission is always emerging from deeper layers in a stellar atmosphere than light in emission lines.
We suspect that there is continuum 3rd light, which is relatively weaker when the amount of stellar light is about twice as strong in the line center than in the continuum. This 3rd light is unlikely light from the wind-wind collision zone because we expect that this region emits  only in emission lines with negligible continuum emission. It might be a stellar member of the $\gamma$\,Vel cluster that is in the field of view of the interferometric measurements {but whose contribution to the total observed luminosity is very small}.

Using a simultaneous fit to all data, the $K$-band brightness difference in the continuum is determined to $K_\mathrm{WR}-K_\mathrm{O}=-0.41\pm 0.07$\,mag, i.e. the WR star is 
${\cal F}_K(\mathrm{WR})/{\cal F}_K(\mathrm{O})=1.46\pm 0.09$ brighter than the O star. As for the diameters, the uncertainty is the standard deviation of the brightness ratios listed in Tab.\,\ref{tab:disk} and not reduced by a factor of two due to the number of measurements.

\section{Astrometric orbit}
\subsection{Fit to AMBER and GRAVITY measurements}

As reported by \citet{Lamberts_etal2017}, there are ten earlier interferometric measurements by AMBER/VLTI, of which we have included eight positions. We have excluded the observation of 7 February 2006, which has too large uncertainties to be of use for an orbital fit. The case of the measurement of 25 December 2004 is discussed in the following. We fitted an orbital solution to the combined twelve observed astrometric positions with the seven free parameters of an orbital solution. As each position is given by two parameters, this is a fit to 24 measurements. We fitted right ascension offsets and declination offsets with the {\em optimize} function of the {\em scipy} library, specifying uncertainties with the parameter {\em sigma} by the corresponding RA- or Dec-sections within the uncertainty ellipses. 

\begin{table*}
\caption[]{Astrometric orbit solutions for $\gamma^2$\,Vel\tablefootmark{a}. Previous solutions are reported by \citet{North_etal2007} (N07) and \citet{Lamberts_etal2017} (L17).
}
\label{tab:orbit}
\centering
\begin{tabular}{c|cc|c|ccc}
\hline%
\noalign{\smallskip}
          & N07 & This paper & L17 & \multicolumn{3}{c}{This paper}\\ 
       & SUSI &  VLTI+SUSI\tablefootmark{b} & AMBER/VLTI & \multicolumn{3}{c}{AMBER/VLTI+GRAVITY/VLTI}\\ 
 $n$ / $f$   &  12 / 1    &      12 / 3  & 9 / 5  & 12 / 7    & 12 / 6 & 12 / 6 \\
\hline
\noalign{\smallskip}
$P$ [d]  & 78.53\tablefootmark{c,g} & $78.5284\pm 0.0006$ & 
            78.53\tablefootmark{c}  & 
         $78.5227\pm 0.0016$ & 78.524\tablefootmark{d} &
         78.526\tablefootmark{d} \\
$T_0$ [MJD]  & $53732.8\pm 0.4$\tablefootmark{f} & 
                  $53732.8$\tablefootmark{b,f} & 
                  $50120.7\pm 0.2$  &
                    $60250.37\pm 0.12$ & $60250.44\pm 0.07$ &
                    $60250.56\pm 0.07$ \\
$a$ [mas]         & $3.57\pm 0.05$ & $3.499\pm 0.008$ & 
                    $3.48\pm 0.02$ & 
                    $3.479\pm 0.008$  & $3.483\pm 0.008$  & 
                    $3.489\pm 0.008$ \\
$b$ [mas]         &                 &              & 
                                    & 
                    $3.296\pm 0.007$  & $3.297\pm 0.007$  & 
                    $3.299\pm 0.007$ \\
$e$         & $0.334\pm 0.003$ & $0.334$\tablefootmark{b} & 
                    $0.326$, 0.334\tablefootmark{e} & 
                    $0.320\pm 0.004$  & $0.322\pm 0.002$ &
                    $0.325\pm 0.002$ \\
$\omega$ [$\deg$] & $67.4\pm0.5$ & $67.4$\tablefootmark{b} & 
                    $68.3\pm 0.9$ & 
                    $67.3\pm 0.4$ & $67.5\pm 0.3$ &
                    $67.8\pm 0.3$ \\
$e \cos(\omega)$  &  &  & & $0.1235\pm 0.0015$ & $0.1232\pm 0.0014$ & $0.1227\pm 0.0014$ \\
$e \sin(\omega)$  &  &  & & $0.2953\pm 0.0039$ & $0.2977\pm 0.0025$ & $0.3014\pm 0.0025$ \\
$\Omega$ $[\deg]$  & $247.7\pm 0.4$ & $247.7$\tablefootmark{b} & $247.6\pm 0.4$ & 
                     $247.0\pm 0.2$  & $247.0\pm 0.2$ &
                     $246.9\pm 0.2$ \\
$i$ $[\deg]$ & $65.5\pm 0.4$ & $65.7\pm 0.2$ & 
                    $64.8\pm 0.7$ & 
                    $64.9\pm 0.3$ & $65.0\pm 0.3$ &
                    $65.1\pm 0.3$ \\
$\chi^2$     &    153\tablefootmark{g} & 57.5 &  & 17.0 & 17.6 & 21.0\\ 
\hline
\end{tabular}   
\tablefoot{
\tablefoottext{a}{The row $n$ / $f$ gives the number of position measurements, $n$, and the number of free fit parameters, $f$. The orbital parameters are: $P$ is the period, $a$ the semi-major axis, $b$ the semi-minor axis, $e$ the eccentricity, $T_0$ the periastron passage in modified Julian date, 
$\omega$ the periastron longitude of the WR star, $\Omega$ the position angle on the sky of the node line, and $i$ the inclination of the orbital plane. There is a relation between axis of the orbital ellipse, $b=a \sqrt{1-e^2}$. The given uncertainties result if either of $a$ or $b$ is used as free parameter in the orbital solution.  
The uncertainties of $e$ and $\omega$ or $e\cos(\omega)$ and $e\sin(\omega)$ result if either pair is used as free parameters.
The sum of normalized uncertainties is $\chi^2 = \Sigma (\delta/\Delta)^2$, where $\delta$ is the deviation of the calculated position from the measurement in right ascension and declination, and $\Delta$ is the estimated measurement uncertainty (1\,$\sigma$, see Tab.\,\ref{tab:GRAVITY_observations}) in the corresponding direction (see text).}\newline
\tablefoottext{b}{Orbit fit to the AMBER+GRAVITY/VLTI observations for fixed $e$, $T_0$, $\omega$, and $\Omega$ adapted from N07.}\newline
\tablefoottext{c}{A period of $P=78.53\pm 0.01$\,d from \citet{Schmutz_etal1997} was adapted by N07 and L17.}\newline
\tablefoottext{d}{Orbit fit for a fixed period (see text).}\newline
\tablefoottext{e}{L17 adapted two eccentricities: 0.326 from S97 and 0.344 from N07.}\newline
\tablefoottext{f}{N07 list $T_0=50120.4\pm 0.4$. The given entry is the periastron passage time at the time of the SUSI observations, using $P=78.53$ to calculate the listed $T_0$.}\newline
\tablefoottext{g}{If the VLTI measurements are compared to positions computed with the N07 orbital solution the listed $\chi^2$ results for an optimized period of $P=78.529$\,d (see text).} 
}
\end{table*}

\begin{figure}
   \centering
   \includegraphics[width=9cm]{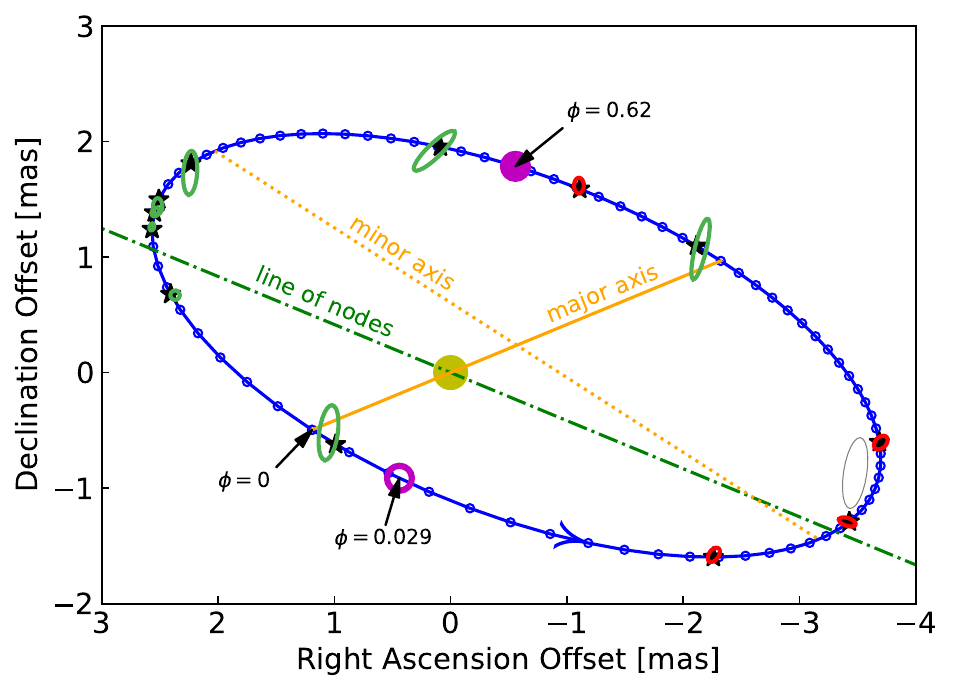}
   \centering
   \caption{Positions of the WR star component relative to the O star component at the focus, which is at coordinates (0,0). The green and red small circles and ellipses represent measured astrometric locations with their dimensions indicating 2.5\,$\sigma$ uncertainties. Locations marked in green are those measured by AMBER/VLTI as reported by \cite[][Tab.\,A1]{Lamberts_etal2017} and those marked in red are measurements by GRAVITY/VLTI as given in Tab.\,\ref{tab:GRAVITY_observations}. The thin gray ellipse marks the uncertainty limit of the position determined by AMBER/VLTI with UT telescopes in December 2004, which has not been used in the present analysis (see text). The black stars mark the calculated positions at the time of the corresponding close-by measurement. The gray dash-dotted line indicates the line of nodes,  
   the yellow line is the major axis of the orbit, the dotted yellow line is the minor axis, and the arrow annotated with $\phi=0$ points to periastron. The blue $\succ$-sign is pointing in the direction of the WR orbit (counterclockwise). The open magenta 
   circle marks the WR star position of closest apparent approach at 2.29\,d (phase 0.029) after periastron with the WR star in the back and the O star in front, whereas the filled-in magenta circle is the position of closest approach at 49.0\,d (phase 0.62) after periastron   when the WR star is in front.
   The blue ellipse shows the calculated apparent orbit and the small open blue dots mark 78 time steps of approximately one day starting and ending at periastron.}
              \label{Fig1}
    \end{figure}

In Tab.\,\ref{tab:orbit} we list astrometric orbit solutions. Each solution is characterized by a value $\chi^2 = \Sigma (\delta/\Delta)^2$, where $\delta$ is the deviation of the measurement of the position from the calculated position in the
right ascension and declination, and $\Delta$ is the estimated uncertainty in the
corresponding direction. The expected value for $\chi^2$ is $N-f$, where $N=2 n$ is the number of measurements with $n$ the number of observed positions, and $f$ is the number of free parameters used to fit the orbit (see 3$^{rd}$ row in Tab.\,\ref{tab:orbit}). 
The combined AMBER/VLTI+GRAVITY/VLTI data set has twelve observed positions thus, $N=24$. A fit to the data set with all seven parameters of an astrometric orbit is expected to have a value of $\chi^2 = 17\pm 5.8$. 
A fit with seven free parameters yields $\chi^2 = 4.7$. Thus, the 7-parameter fit yields a $\chi^2$ value that is more than $2\sigma$ smaller than expected and we suspect that the uncertainties 
have been overestimated. We therefore decrease all uncertainties by a factor of 1.86 in order to get $\chi^2$ values that yield  the expected sum. This does not change the solutions but yields smaller uncertainty estimated by the covariance matrix returned by the optimize routine as listed in table Tab.\,\ref{tab:orbit}. 

In Fig.\,\ref{Fig1} we illustrate the relative apparent orbit of the WR companion around the O star. The measurements are represented by uncertainty ellipses and it can be seen that all measurements except one fit the calculated orbital solution. The exception is the first astrometric observation of $\gamma^2$\,Vel in December 2004 by ESO's AMBER/VLTI with three UT telescopes as reported by \citet{Milour_etal2007}. In Fig.\,\ref{Fig1} (see enlargements around the measurements in Fig.\,A1) it is apparent that the measured separation is too small, and a comparison to the computed position reveals that the right ascension offset deviates by $0.18$\,mas, which is 1.8\,$\sigma$. The declination offset differs by $0.27$\,mas from the calculated position but because of the larger measurement uncertainty this corresponds only to 1\,$\sigma$ (for the measurement uncertainty see Tab.\,A1 of \citet{Lamberts_etal2017}). 
There is an added difficulty in that the reported right ascension offset given in \citet{Milour_etal2007} deviates from the value given by \citet{Lamberts_etal2017}.  In fact, \citet{Milour_etal2007} have described difficulties during the observations in their Section 2.2 and it is possible that later corrections have been applied. The observation is not an outlier in a scientific sense, as it does not deviate by more than 2\,$\sigma$, but we feel it is justified to exclude the December 2004 AMBER/VLTI position from our analysis. 

The uncertainties of the fitted parameters are computed from the diagonal of the covariance matrix, which is returned by the {\em optimize} routine. These uncertainty estimates do not include co-dependencies, which are present for this data set: The resulting parameters $a$, $e$, $\omega$, and $T_0$ are systematically sensitive to the period. In addition to the best fitting orbit solution, which yields a period of $78.5227\pm 0.0016$ we also list  orbit solutions with a fixed period of $P=78.524$\,d and 78.526\,d. It can be seen that with longer periods the four dependent parameters all change to larger values, whereas $\chi^2$ is not significantly increased. This is in contrast to varying only one parameter by 1\,$\sigma$, which yields $\chi^2\approx 23$.

We conclude that by including co-dependencies the uncertainty of the period is twice as large as determined from the covariance matrix: $P=78.523\pm 0.004$\,d.

\subsection{Comparison with the SUSI solution}
\citet{North_etal2007} have observed the astrometric orbit of $\gamma^2$\,Vel with the intensity interferometer SUSI (Sydney University Stellar Interferometer; \citet{Davis_etal1999}). They measured the variations in squared visibility during 20 nights with a baseline of 80\,m. Unfortunately, they did not calculate nightly apparent separation and position angle from their data but modeled the orbit directly from their combined visibility data set. Therefore, comparing with their observations is not straight forward, as we only have their observation dates and their orbital solution. We reconstruct their not-given nightly positions by calculating the positions for the SUSI observation dates with the elements of their orbital solution.

\begin{figure}
   \centering
   \includegraphics[width=9cm]{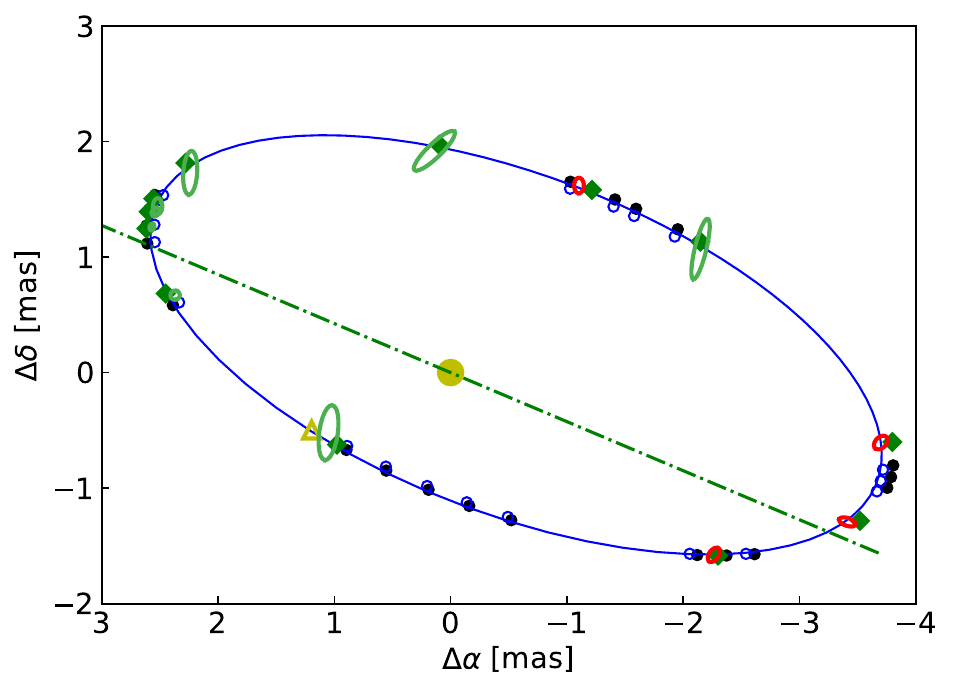}
   \centering
   \caption{Reconstructed positions of the measurements with the intensity interferometer SUSI by \citet{North_etal2007} (black filled circles) compared to the  
   orbit resulting from the best VLTI solution (blue line) as well as the positions calculated 
   with the angular information from \citet{North_etal2007} and fitted VLTI positions (open blue circles), which is the second column in Tab.\,\ref{tab:orbit}. The small green ellipses mark the measurements and uncertainties of the VLTI measurements and the green diamonds denote the positions calculated using the orbital solution of \citet{North_etal2007}.} 
              \label{Fig:SUSI}%
    \end{figure}
Fig.\,\ref{Fig:SUSI} compares the reconstructed positions with our best astrometric solution with $P=78.5227$\,d (see Tab.\,\ref{tab:orbit}). 
The reconstructed SUSI measurements are very close to our apparent orbit solution except for phases around 0.3, for which the SUSI points are at significantly larger separations. There is also a deviation in angle, which is most apparent close to periastron. It is clear that the period adopted by \citet{North_etal2007}, $P=78.53$, is not compatible with the VLTI observations (see Sect.\,4). Therefore, we optimized the period to match the VLTI measurements while keeping all other parameters of the SUSI orbit. This yields better angle agreement, but does not fully remove the systematic differences. When only keeping the angle information of the SUSI orbit, that is the parameters $e$, $T_0$, $\omega$, and $\Omega$ are fixed and $P$, $a$, and $i$ are fitted, we find a solution that is
compatible with the VLTI observations (see Tab.\,\ref{tab:orbit}) but it still has systematic deviations in angle, in particular with the AMBER/VLTI measurement closest to periastron and GRAVITY/VLTI close to apastron, which deviate by more than $6\sigma$. In the sum of deviations, $\chi^2 = \Sigma (\delta/\Delta)^2=57.5$, the solution with the 4 SUSI orbital parameters and 3 free parameters deviates by $5.6\,\sigma$ from the expected value of $\chi^2_\mathrm{exp} = 21\pm 6.5$. 

We are of the opinion that the SUSI observations have systematic deviations with respect to the VLTI measurements. This is most apparent in the difference in separation but is also present in angle. Our hypothesis is that this is due to the influence of emission from the wind-wind collision zone, which has strong emissions in the red filter used by the SUSI instrument, including the H$\alpha$ line (see Section 4 of \citet{Richardson_etal2017}. This could have offset the apparent position of the O star.

\section{Orbital period}
\citet{Schmutz_etal1997} have investigated the orbital period in their Section 3 by comparing their spectroscopic observations of 1997 with the 1919 data set of \citet{Perrine1920} and with the 1904/1913 data of the Mills Expedition to Chile re-measured by \citet{Pike_etal1983}. They obtained periods of $78.522\pm 0.005$\,d and $78.535\pm 0.01$\,d, respectively. Given the uncertainties quoted, both results agree with our period determination in Sect.\,3.1. However, in order to break the co-dependencies of the orbital parameters a more accurate period determination is needed. Since 1997 \citet{Richardson_etal2017} have published new spectroscopic orbital velocity-shift measurements, and there are also additional older published measurements of velocity variations, which can be used for period estimates.

In Tab.\,B.1 we list periods derived with two methods. The two values agree within their uncertainties, which enhances the probability that the average period is a reliable estimate of the period. However, we are aware that the derived periods are subject to systematic influences. Thus, it is not justified to use statistical methods to derive the uncertainty of the best estimate, which would be $\sigma=0.0005$\,d. Instead, we are of the opinion that it is justified to cover the majority of the periods derived in Appendix B, and give as best estimate $78.524\pm 0.002$\,d.
The ephemeris for periastron dates at $\omega=67.5^\circ$ (Tab.\,\ref{tab:orbit} ) is
   \begin{equation}
   \label{eq:1}
      T_0(E)=\mathrm{BJD}\,2460250.94 \pm 0.1 + (78.524\pm 0.002)\,\mathrm{d} \times E \,.
   \end{equation}

\section{Distance to the $\gamma^2$\,Vel system\label{Sec:5}}
In order to convert the astrometric angle measurements into length we need the radial velocity measurements. 
The most recent data set of spectroscopic observations covering a whole orbit is by \citet{Richardson_etal2017}. In their publication they have concentrated on understanding the contribution from the wind-wind collision zone to the emissions. But they did not disentangle the stellar radial velocity measurements from systematic influences. Therefore, the radial velocities reported by \citet{Schmutz_etal1997} are still the best choice because they have removed the influence of the WR emission on the measured O star absorption.

With the combined velocity amplitude $K=160\pm 3$\,km/s \citep{Schmutz_etal1997} and the measured angle separation of the major axis given in Tab.\,\ref{tab:orbit}, $a=3.483\pm 0.008$\,mas, we derive a distance $d=346\pm 7$\,pc. This distance is slightly larger than the values previously reported by \citet{North_etal2007} and \citet{Lamberts_etal2017}, but not significantly. However, the present distance is 6\,\% smaller than the distance reported by \citet{Milour_etal2007} because their value is based on the AMBER/VLTI measurement in December 2004, which is a measurement that we excluded in Sect.\,3.1 above.

\citet{MaizApellaniz_etal2022} have investigated the stars in the $\gamma$\,Vel cluster Villafranca O-024 and found 138 members at a distance of $d_\mathrm{cluster}=336\pm 1$\,pc. The VLTI distance is 10\,pc or 1.6\,$\sigma$ larger than the distance of the cluster. Formally, with this normalized error $\gamma^2$\,Vel would not be included in the filtered members of the cluster as \citet{MaizApellaniz_etal2022} applied a 1.4\,$\sigma$ criterion. However, since the distance deviation is statistically not significant we infer that $\gamma^2$\,Vel is a member of the $\gamma$\,Vel cluster and conclude that the cluster distance represents the most accurate distance. This conclusion is supported by Hipparcos astrometry \citep{ESA1997}  $\gamma^2$\,Vel, which yielded $d_\mathrm{Hip}=258\pm 35$\,pc, $\mu\,\alpha_\mathrm{Hip}=-5.93\pm 0.53$\,mas/a, and $\mu\,\delta_\mathrm{Hip}=9.90\pm 0.43$\,mas/a. These values agree well in proper motions with the cluster values as measured by GAIA \citep{MaizApellaniz_etal2022}. As the uncertainty of the VLTI astrometric distance derived above is dominated by the uncertainty of the 
amplitude in the radial velocity curve, we estimate
that the correct velocity amplitude should   be $K_\mathrm{est}=155.6\pm 0.6$\,km/s, by using the cluster distance and the measured angle separation of the stars.

\begin{table}
\caption[]{System and stellar parameters of $\gamma^2$\,Vel\tablefootmark{a}. The stellar parameters are rescaled results of \citet{deMarco_etalI1999} (DM99) and \citet{deMarco_etalII2000} (DM00).
}
\label{tab:system}
\centering
\begin{tabular}{ccl} 
$P$ [d]  & $78.524\pm 0.002$ &  \\
$a$ [mas]         & $3.483\pm 0.008$ &  \\
$e$         & $0.322\pm 0.004$ \\
$T_0$ [MJD]  & $60250.44\pm 0.07$ \\
$\omega$ [$\deg$] & $67.5\pm 0.4$ \\ 
$\Omega$ $[\deg]$  & $247.0\pm 0.2$ \\ 
$i$ $[\deg]$ & $65.0\pm 0.3$ \\
\noalign{\smallskip}
$d$ [pc] & $336\pm 1$ & MA22\\
\noalign{\smallskip}
$a$ [AU] & $1.170\pm 0.004$ & \\
$M_\mathrm{sys}$ [M$_\mathrm{\odot}$] & $34.7\pm 0.4$ \\
\noalign{\smallskip}
$q$ & $0.315\pm 0.022$ & S97\\
$M_\mathrm{O}$ [M$_\odot$] & $26.4\pm 0.5$ \\
$M_\mathrm{WR}$ [M$_\odot$] & $8.3\pm 0.7$ \\
\noalign{\smallskip}
$K(\mathrm{WR})-K(\mathrm{O})$ [mag] & $-0.41\pm 0.07$ & \\
$K_s(\mathrm{O})$ [mag] & $3.08\pm 0.8$ & \\
$R_\mathrm{O}$ [R$_\odot$] & $13.5\pm 1.0$ & \\
$R_\mathrm{O}$/$R_\mathrm{WR}$ & $4.0\pm 0.5$ & DM00\\
$R_\mathrm{WR}$ [R$_\odot$] & $3.4\pm 0.5$ & \\ 
$T_\mathrm{eff}(\mathrm{O})$ [K] & $35000\pm 300$ & DM99 \\ 
$T_*(\mathrm{WR})$ [K] & $57100\pm 1700$ & DM00 \\ 
$L_\mathrm{O}$ [L$_\odot$] & $(2.5\pm 0.4)\times 10^5$ & \\ 
$L_\mathrm{WR}$ [L$_\odot$] & $(1.1\pm 0.4)\times 10^5$ & \\
$V_\mathrm{rot}^\mathrm{O} \sin\,i$ [km/s]& $220\pm 20$ & B90\\
$V_\mathrm{rot}^\mathrm{O}$ [km/s]& $240\pm 20$ & \\
\end{tabular}   
\tablefoot{
\tablefoottext{a}{The orbital period $P$ is determined in Sec.\,4. The orbital parameters $a$, $e$, $T_0$, $\Omega$, and $i$ are taken from the second last column in Tab.\,\ref{tab:orbit}  for the adopted period. The distance $d$ is the cluster distance reported by \citet{MaizApellaniz_etal2022} (MA22). The semi major axis in astronomical units is derived from the angular separation and the distance. The system mass $M_\mathrm{sys}$ results from Kepler's third law using $a$ and $P$. The ratio of the masses, $q$, is from \citet{Schmutz_etal1997} (S97). The radius of the O star results from the brightness of the O star in the Ks band. The component's radii ratio is from \citet{deMarco_etalII2000} (DM00) and the WR star's radius results from this ratio and the O star radius.
The effective temperature of the O star was determined by \citet{deMarco_etalI1999} (DM99) and the temperature at the inner boundary of the adopted WR wind structure, $T_*$, by DM00. 
The O star rotational velocity was measured by \citet{Baade_etal1990} (B90).
}\newline
}
\end{table}

\section{Discussion\label{Diskusion}}
\subsection{X-ray variations} 

\begin{figure}
   \centering
   \includegraphics[width=9cm]{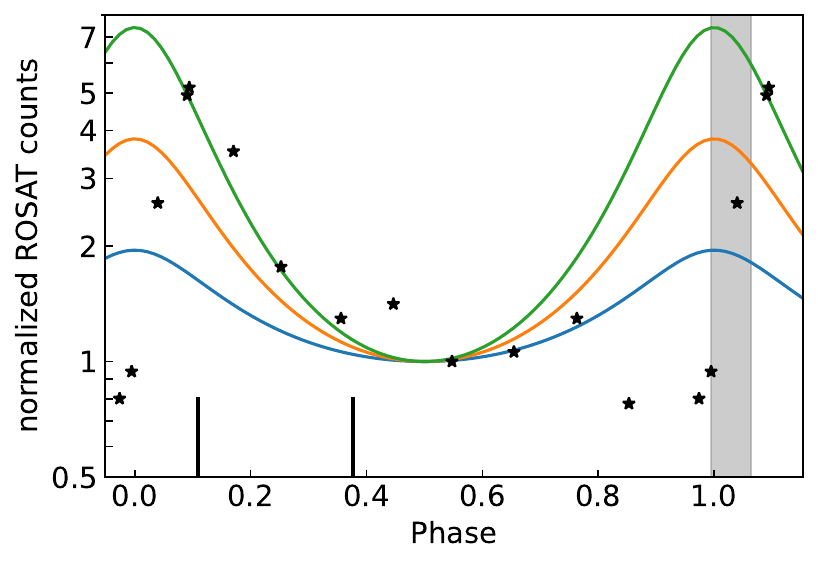}
   \centering
   \caption{ROSAT PSPC X-ray counts (black stars) from \citet{Willis_etal1995} normalized to the counts at apastron (phase 0.5).  
   The blue curve shows the expected behavior if the intensity had an inverse separation dependence, also normalized to apastron, the green line is the squared inverse separation, and the red line denotes the cubed inverse separation. 
   The two vertical black lines at the bottom of the figure mark the phases of the XMM observations analyzed by \citet{Schild_etal2004}.
   The gray vertical area from phase 0.995 to 1.064 marks the length of the eclipse of the stagnation point of the wind-wind collision by the O star disk (see Fig.\,\ref{Fig:Xloc}).
   }
              \label{Fig:Xray}
    \end{figure}

\begin{figure}
   \centering
   \includegraphics[width=9cm]{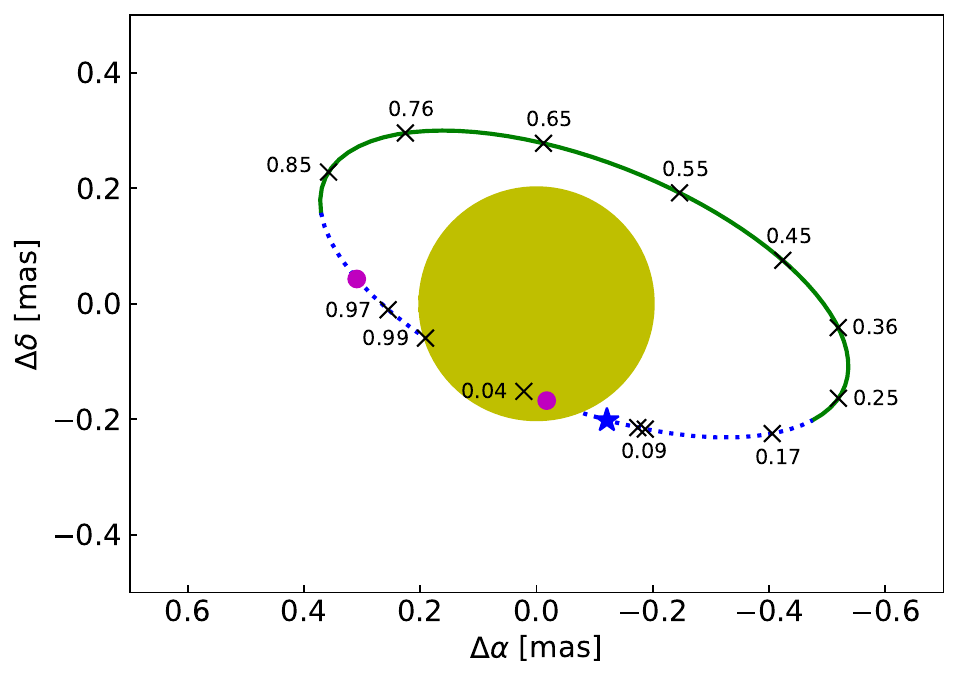}
   \centering
   \caption{Estimated apparent locations of the stagnation point of the wind-wind collision, relative to the position of the O star in the center. The yellow circle has the size of the O star. The locations labeled with an orbital phase are ROSAT observation (black crosses) as plotted in Fig.\,\ref{Fig:Xray}. 
   The blue dotted line marks locations behind the O star; the green line gives locations in front of the O star. 
   The blue star denotes the phase 0.08 of the CHANDRA observation \citep{Skinner_etal2001} and the two magenta circles mark the phases 0.95 and 0.05 of the ASCA measurements \citep{Stevens_etal1996}. The stagnation point locations are assumed to be on the line connecting to the WR star positions and a fraction of 6.9 of the separation away from O star (see text). 
    }
              \label{Fig:Xloc}
    \end{figure}

\citet{Willis_etal1995} observed $\gamma$ Vel with {\it ROSAT} from 1991 December to 1993 May on 13 occasions in the 0.1 to 2.5 keV energy range.  They found a light curve with a pronounced maximum ($\sim$0.43 cts/s) right after periastron, followed by a decline to a first minimum ($\sim$0.10 cts/s) at apastron and then a deeper minimum ($\sim$0.05 cts/s) between phases 0.9 and 1. We reproduce these data in  Fig.\,\ref{Fig:Xray}, normalized to the value at apastron.  Interestingly,  \citet{Willis_etal1995} also showed that the steep increase takes place only in ROSAT's hard X-ray band ($\ge \sim$1 keV, concluding that these hard X-rays are superposed on a softer, relatively constant kT=0.19 keV component associated with the WR wind.  They also concluded that the hard X-rays  originate in the wind-wind collision, but cannot be observed at orbital phases in which they  have to traverse the WR wind to reach us.  This is, however, not the only process affecting the X-ray light curve.

Fig.\,\ref{Fig:Xloc} is a plot showing the location of the stagnation point of the wind-wind collision.  This is the point at which the wind momentum of the WR star is equal to that of the O star, and is calculated with their mass-loss rates and terminal wind speeds.  This figure shows that the stagnation point is eclipsed by the O-star disk during the phase interval $\sim$0.97 - 1.08.  Thus, the fraction of X-ray emission originating around the stagnation point disappears from the line-of-sight during this phase interval.  In addition, any uneclipsed X-rays are likely to suffer from enhanced absorption as they traverse the leading edge of the wind-wind collision, which is predicted to have an over-density due to the interaction  (see, e.g.\, Fig.\,12 of \citet{Henley_etal2005} and Fig.\,9 of \citet{Schild_etal2004} for sketches of the geometry).

The dependence on orbital separation of the X-ray intensity shown in Fig.\,\ref{Fig:Xray} is compared to three functions that depend only on distance.  The smooth decline after periastron follows a $F_X \propto s^{-3}$ law, where $s$ is the separation between the two stars.  This, however, does not necessarily imply that the separation is the dominant factor affecting the X-ray brightness. As noted above,  the observed modulation of the ROSAT high-energy X-rays seems to be largely a consequence of a varying opacity along the line of sight and the occultation of the stagnation point by the O-star disk.   

Results from modeling the X-ray absorption along the line of sight to the source of hard X-rays are lacking because of the scant X-ray spectral information that exists. Spectral fitting only seems to be available for two XMM-Newton observations obtained at phases 0.11 and 0.37 \citep{Schild_etal2004}.   These phases are  indicated by the two black lines in Fig.\,\ref{Fig:Xray}. The analysis performed by \citet{Schild_etal2004} of the column density of the WR wind toward the X-ray source at orbital phase 0.37 led to the conclusion that the WR wind is clumped, which significantly complicates a broader analysis of the line-of-sight opacities.

Given the above, our conclusion for this section is that without a detailed modeling of the WR wind opacity and the density in the wind-wind collision leading and trailing arms, it is not possible at this time to address the dependence of X-ray emission on orbital separation.

\subsection{Radial velocity curves and perturbations from the wind-wind collision}

In Table \ref{radvel} we list the lines from Schmutz et al. (1997) and the derived orbital parameters from fits to their RV  curves.  The individual lines yield eccentricities in the range 0.31-0.36.
The fits to RV curves performed by Richardson et al. (2017) yielded eccentricities in the range 0.11 to 0.42. They speculated that these differences were likely due to wind-wind collision effects.  The presence of narrow sub-peaks superposed on the broad WR lines is often interpreted as arising in the WCR with models of the expected sub-peak velocity as a function of orbital phase constructed based on simplifying approximations \citep{Luehrs_1997, 2000MNRAS.318..402H, 2009MNRAS.395..962I, 2021MNRAS.503..643W}.  However, the manner in which these sub-peaks affect the overall RV curve is difficult to assess because the flux of optical line emission from the WCR, though likely small compared to that of the WR wind, is not known.    In addition, wind-wind collisions are only one of the processes that can produce systematic line profile variations and the perturbations of RV curves to the various effects is likely to depend on each particular binary system's parameters.  Hence, it is not possible at this time to determine which lines are best for orbital  determinations in general.  

For the particular case of $\gamma$ Vel, Table \ref{radvel} indicates that the lines of 
\ion{He}{ii}\,$\lambda$\,4686 and 
\ion{C}{iv}\,$\lambda\lambda$ 4441, 4786, and 7730 yield radial velocity curves that best represent the orbital motion (Fig.\,\ref{Fig:C2} and \ref{Fig:C3}). 
These are all lines emerging from the innermost ionization structure of the WR wind. The lines that deviate from the predicted radial velocity curve are lines from the recombined region of the wind. 
\citet{DeMarco2002} demonstrated that \ion{C}{ii} emission lines deviate even more clearly from the WR radial velocity curve.
The distortion effects on the measured radial velocities are a shift of the phase and an enlargement of the velocity amplitudes (see Fig.\,\ref{Fig:C1}).

\subsection{Stellar parameters\label{stellar_parameters}} 

\citet{deMarco_etalII2000}  derived the stellar parameters of $\gamma^2$\,Vel using the Hipparcos distance $d_\mathrm{Hip}=258
\pm 35$\,pc, which implies that we need to account for a 0.6\,mag difference of the distance modulus relative to the adopted cluster distance. Thus, there is a need to rescale the stellar parameters to the presently best cluster distance. 
In addition, in Sect.\,2 we determined a brightness ratio of the components in the K-band that allows to determine the stellar radii without the uncertainty of the reddening.

The most precise infrared brightness measurements of $\gamma^2$\,Vel are reported by \citet{Kimeswenger_etal2004}, with $J=2.178\pm 0.057$\,mag and $K_s=2.004\pm 0.032$\, mag. 
The flux contribution of emission lines to the measured flux in the $K_s$-band filter is $9.0\pm 1.5$\,\%, which we evaluated from the normalized GRAVITY K-band spectra multiplied by the transmission curve of the $K_s$ band filter obtained from the web site of the SVO Filter Profile Service\footnote{http://svo2.cab.inta-csic.es/theory/fps3/}. This yields a continuum brightness of the $\gamma^2$\,Vel system of $K_\mathrm{s\ cont.}=2.098\pm 0.046$\,mag.

With the WR star to O star brightness difference in the continuum of -0.41\,mag reported in Sect.\,2 this yields $K_\mathrm{s\ cont.}(\mathrm{WR})=2.665\pm 0.067$\,mag and $K_s(\mathrm{O})=3.075\pm 0.076$\, mag, where the uncertainty is the root of the added uncertainties squared of the system brightness and the components ratios to the system brightness.

The $K_s$-band zero point flux provided by 
SVO Filter Profile Service
yields a K$_s$-band flux of ${\cal F}_{Ks}=(6.72\pm 0.19)\times 10^{-14}$\,W/m$^2$/nm 
and using the emission line correction and the component's brightnesses given above yields 
${\cal F}_{Ks}(\mathrm{O})=(2.51\pm 0.15)\times 10^{-14}$\,W/m$^2$/nm and ${\cal F}_\mathrm{{\it Ks}\ cont}(\mathrm{WR})=3.66\pm 0.22)\times 10^{-14}$\,W/m$^2$/nm.

From the observed flux divided by a model atmosphere flux we get the apparent angle of the stellar diameters, which are $\Theta_\mathrm{flux}$(O)$=0.37$\,mas using the atmosphere flux from the O star model of \citet{deMarco_etalI1999} and $\Theta_\mathrm{flux}$(WR continuum)$=0.12$\,mas using the WC star atmosphere flux of \citet{deMarco_etalII2000}. With the cluster distance of $d_\mathrm{cluster}=336\pm 1$\,pc we infer the stellar radii 
$R_\mathrm{flux}$(O)$=13.5\pm 1.0$\,R$_\odot$ and $R_\mathrm{flux}$(WR cont.)$=4.2\pm 0.3$\,R$_\odot$, where the uncertainties account for contributions from brightness, temperature, and distance. The uncertainty of the WR star does not include the suspected systematic problem of the WR model atmosphere in the infrared discussed below.

\begin{figure}
   \centering
   \includegraphics[width=9cm]{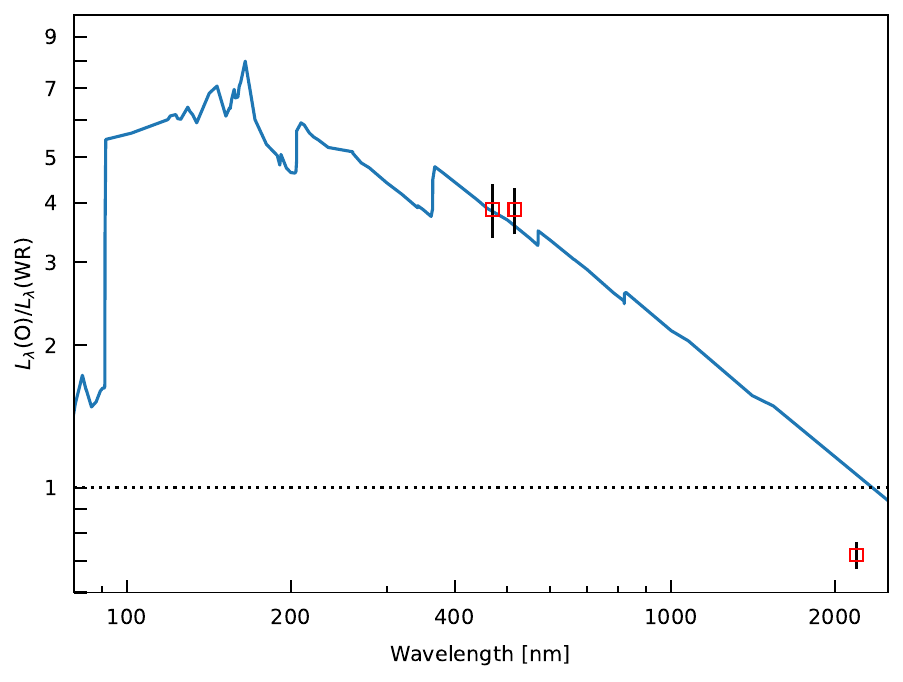}
   \centering
   \caption{Luminosity-ratio of the O7 star to the WR star. This figure shows the same ratio as Fig.\,1 of \citep{deMarco_etalII2000} but for a larger wavelength range including the K-band. The red squares with uncertainties are the ratios $L_\lambda(\mathrm{O}/L_\lambda(\mathrm{WR})=3.88^{+0.5}_{-0.4}$ at $\lambda=470$\,nm \citep{deMarco_etalII2000}, $3.87\pm 0.44$ at $\lambda=516$\,nm \citep{deMarco_etalI1999}, and $0.69\pm 0.03$ at $\lambda=2160$\,nm (Tab.\,2).
   }
              \label{Fig:lum_ratio}
    \end{figure}

Compared to radii obtained when scaled to the new distance we find that the radius of the O star is smaller and the radius of the WR star is larger. The reason for this is that we find a disagreement between the observed flux ratios in the optical and in the infrared compared to the model luminosity ratios. The stellar atmosphere models in the cited publication yield a luminosity ratio of the two stars as given by \citet{deMarco_etalII2000} in their Fig.\,1. In Fig.\ref{Fig:lum_ratio} we repeat their figure but with enlarged wavelength coverage including the Ks-band. It is evident that the model luminosity-ratio does not fit at all the observed infrared flux ratio. 

It could be argued that the source of the discrepancy is in the flux ratio of the WR to the O star that was inferred from the interferometry in Sect.\,\ref{Sec:2}. However, the problem is so large that possible systematic influences of a not fully adequate two-disk model, which has been used to deduce the flux ratio, are outside of what we think could be feasible. Also, the O star's atmosphere model 
cannot be wrong by the amount seen in Fig.\ref{Fig:lum_ratio} because the NLTE-effects make its flux distribution deviate only moderately from a Kurucz atmosphere with the same temperature and gravity.  
Therefore, most likely, the problem is the flux distribution of the WR star model in the infrared, which should be less steep than computed. 

We investigated whether other WR atmosphere models reproduce better the WR continuum flux distribution. \citet{Lamberts_etal2017} also used a {\sc cmfgen} model \citep{Dessart_etal2000}
to calculate the theoretical spectrum of the WR star in $\gamma^2$ Vel as \citet{deMarco_etalII2000}, which we used for Fig.\,\ref{Fig:lum_ratio}. The \citet{Lamberts_etal2017} model was provided to us by \citet{Dessart2025}. We found that the two models have essentially the same flux continuum distribution from the optical to the infrared. We also compared to the publicly available Potsdam Wolf-Rayet Models, {PoWR}\footnote{Available at https://www.astro.physik.uni-potsdam.de/$\sim$wrh/PoWR} \citep{Sander_etal2012}, 
and found that if the line strengths fit the observations then the continuum flux distribution also agrees with the {\sc cmfgen} modes. Thus, there is no WR atmosphere available to us that could reproduce the observed flux continuum between the optical and the K-band.

If we assume that the WR atmosphere model is 
correct in the optical wavelength region then we can 
adopt from the solution of \citet{deMarco_etalII2000} the radius 
ratio\footnote{The entry $R(\mathrm{O})=12.4$\,R$_\odot$ in Tab.\,1 of \citet{deMarco_etalII2000} should read $R(\mathrm{O})=12.63$\,R$_\odot$ to reflect a correction of -0.04\,mag in absolute O star magnitude as explained in their Sect.\,3.}
$R(\mathrm[O])/R(\mathrm[WR])=4.0\pm 0.5$, which reproduces the flux ratios of the components in the optical. Therefore, we calculate from the O star's radius derived above from the observed $K_s(\mathrm{O})$ brightness the radius of the WR star using the radius ratio. This yields 
$R$(WR)$=3.4\pm 0.5$\,R$_\odot$.

The stellar effective temperatures are, to first order, not affected by different brightness ratios, because the temperature is determined by ratios of line strengths. Thus, we adopt for the effective temperature of the O star 
$T_\mathrm{eff}(\mathrm{O})=35,000\pm 300$\,K \citep{deMarco_etalI1999} and $T_*(\mathrm{WR})=57,100\pm 1700\,$K \citep{deMarco_etalII2000} for the WR star. The temperatures and radii yield the luminosities 
$L(\mathrm{O})=(2.46\pm 0.42)\times 10^5$\,L$_\odot$ and  
$L(\mathrm{WR})=(1.1\pm 0.4)\times 10^5$\,L$_\odot$.

The determined luminosities are smaller than those given by \citet{2024MNRAS.528.2026C}. The main reason is reddening. We have determined the radius of the O star from the Ks-band measurements without a correction for reddening. The spectral energy distribution of the O star from the K-band to visual magnitudes yields a reddening of $E(B-V)=0.02\pm 0.02$\,mag, which would imply a 0.7\,\% correction of the Ks magnitudes, which is much smaller than the uncertainty of the observed brightness.  \citet{2024MNRAS.528.2026C} have applied a reddening correction of $E(B-V)=0.09$\,mag. The problem of the incorrect O star to WR star ratio revealed in Fig.\,\ref{Fig:lum_ratio} has been compensated by applying a higher reddening. \citet{deMarco_etalII2000} have commented on contradictory reddening values obtained from $ubv$ photometry and from the observed energy distribution compared to the model energy distribution. With our low reddening value, this contradiction has been resolved.

There are two other WR+O systems, WR137 and WR138 that have been resolved by interferometry \citep{Richardson_etal2016, Richardson_etal2024, Holdsworth_etal2024}.
These two systems are highly reddened and because the reddening value may compensate for a incorrect WR continuum energy distribution from the optical to the infrared it is not possible to judge weather there is a similar problem in these systems. However, there is one case for which it is certain that there is clear problem with the K-band diameter of the model: The eclipsing binary WR 139 \citep{Kurosawa_etal2002}. The K band eclipse was observed wider than their model could explain, but otherwise the light curves in the optical have been fitted extremely well.

\subsection{Evolutionary stage}
\begin{figure}
   \centering
   \includegraphics[width=9cm]{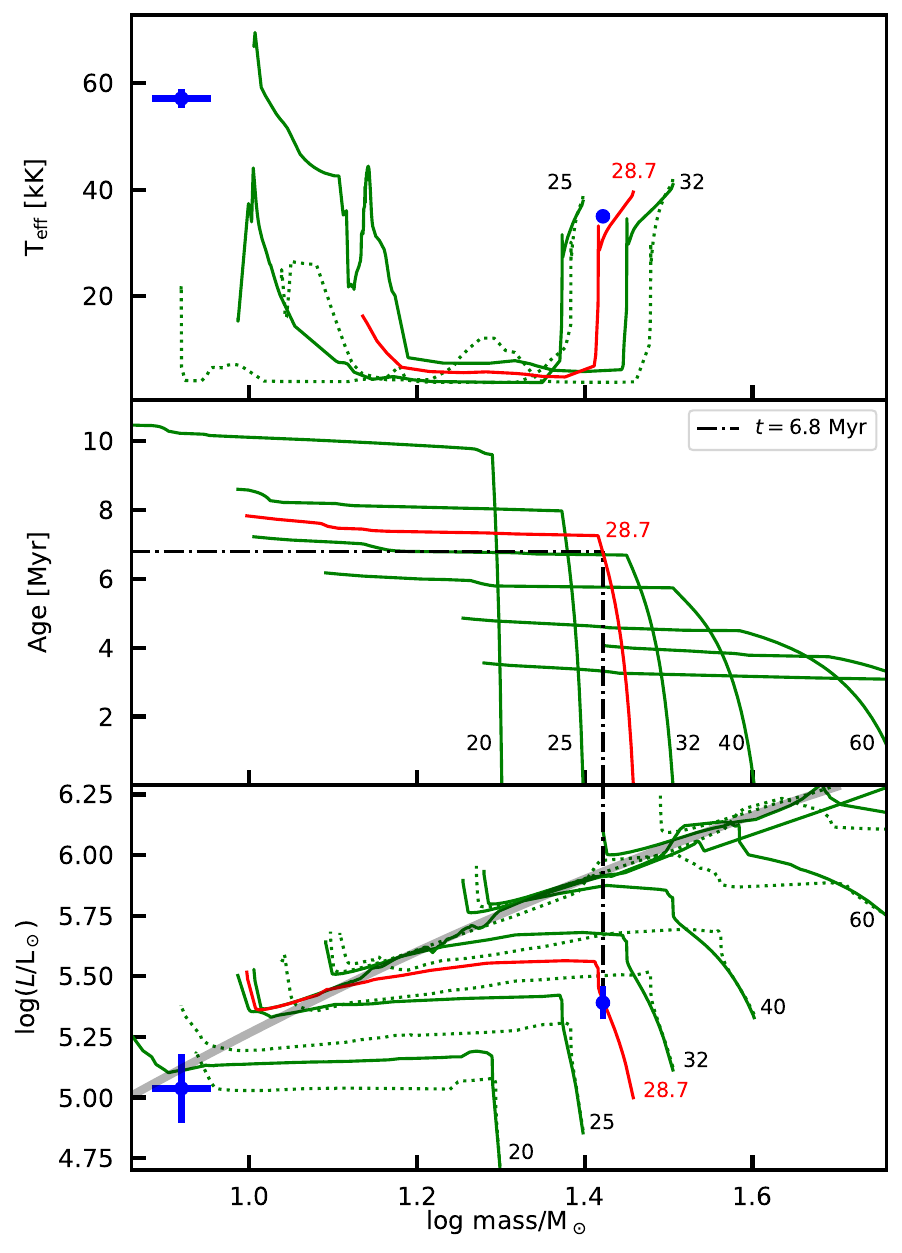}
   \centering
   \caption{Evolutionary tracks at solar metallicity (Z=0.014) from \citet{Ekstrom_etal2012} without rotational mixing (green dotted) and including rotational mixing (green lines). The labels denote the initial mass.
   Also plotted is 
   an interpolated track at solar metallicity and rotationally mixing for $M_\mathrm{ini}=28.8$\,M$_\odot$ (red line).\newline
   Bottom panel: Mass-Luminosity diagram. The gray curve is the quadratic mass-luminosity relation for WR stars given by \citet{SchaererMaeder1992}. The locations of the WR star (left) and the O star (right) in this diagram are marked with blue dots and error bars. \newline
   Middle panel: Mass-Age diagram. The age of the O star is estimated from the 28.8\,M$_\odot$ track (red line) yielding an age of 6.9\,Myr. 
   \newline
   Top panel: Mass-Temperature diagram. The locations of the WR star (left) and the O star (right) in this diagram are marked with blue dots and error bars.  
   }
              \label{Fig:track}
    \end{figure}

In this section we discuss evolutionary models that may reproduce  $\gamma^2$ Vel's fundamental parameters.  We focus first on the O7-star.  

If we place the O7-star's temperature and luminosity (Tab.\,\ref{tab:system}) on the Hertzsprung-Russell diagram and compare with the evolutionary track of \citet{Ekstrom_etal2012}, we find that an interpolated track
for initial mass $M_\mathrm{ini}=31.3$\,M$_\odot$ passes through the observed data point.
However, the model mass at this evolutionary stage is $M=29.3\,\mathrm{M}_\odot$, which is 2.9\,$\mathrm{M}_\odot$ larger than the value determined from observations.
This is the well-known "mass-discrepancy problem"  \citep{Herrero_etal1992},
in which the dynamically determined masses are systematically found to be smaller than the evolutionary masses.  \citet{2012ApJ...748...96M}   showed that dynamical masses  are $\sim$12\% smaller than evolutionary masses or, alternatively, 0.2 dex brighter, which is in good agreement with the mass difference found for the O star.

An alternative approach to the HRD is presented in Fig.\,\ref{Fig:track}, where we plot the  effective temperature (top), age (middle), and  luminosity (bottom) as a function of the corresponding stellar given in the models.  We chose the stellar mass for the abscissa because it is the most strongly constrained of the O7-star's parameters and it is independent of assumptions.  We analyze the stellar evolution models from \citet{Ekstrom_etal2012} for solar metallicity and in Fig.\,\ref{Fig:track} we show tracks corresponding to both rotationally mixed and  non-rotating models.  The bottom panel of this figure shows that 
the observed luminosity constrains the tracks to those with an initial mass M$_{init}$ between 25 and 50 M$_\odot$.  However, only an interpolated, rotationally mixed M$_{init}$=28.7 M$_\odot$ model is consistent with the observed current stellar mass. The effective temperature at this evolutionary stage is T$_\mathrm{eff}\approx 31'000$\,K (top panel), cooler than the 35,000\,K that was determined from the photospheric absorption lines by \citet{deMarco_etalI1999}. This $\sim$10\% difference reflects the mass discrepancy problem noted above.

From the middle panel of Fig.\,\ref{Fig:track} we find that at an age of 6.8\,Myr the $M_\mathrm{ini}=28.7$\,M$_\odot$ track has reached the observed mass of the O star.  Also, at this age a model with M$_{init}$=  32\,M$_\odot$ has become a $\sim$10\,M$_\odot$ WR. This would be consistent with the more massive component of the binary having reached the WR phase first.  However, it also means that this star lost $\sim$24\,M$_\odot$ in the process.  With the above assumption that the original mass of the O7-star was 28.7\,M$_\odot$, most of the mass lost to the WR progenitor would have been lost to the system.  Upon first analysis, this seems strange because the radius during the red supergiant phase of a 32\,M$_\odot$ star is significantly larger than the current orbital separation ($a$=252\,R$_\odot$) and a Roche Lobe Overflow (RLO) mass-transfer phase might be expected to have occurred.  
However, \citet{2009MNRAS.400L..20E} reached a similar conclusion based on the comparison of $\gamma$ Vel's parameters with a grid of binary evolution models. This author notes that the system is likely the product of post-main-sequence mass transfer (Case B), which  occurs on a thermal time-scale, which is short and does not lead to significant accretion onto a companion.  This scenario is favoured also because tidal forces have less time to circularize the orbit and the resulting binary remains eccentric.

\citet{2024MNRAS.528.2026C}   assumed WR progenitor had 35\,M$_\odot$, based on the measured oxygen abundance in the current outer WR wind region.  The determinations of abundances in the O7-star by \citet{2024MNRAS.528.2026C} indicates no excess He or N, as might be expected had it accreted a large fraction of the WR progenitor's outer layers, hence supporting the idea that negligible accretion from the WR progenitor onto the companion occurred.

A single star evolution model with an initial mass of $\sim20\,\mathrm{M}_\odot$ could reach the observed parameters of the WR star but only after 10\,Myr. Thus, because the O7-star cannot be older than $\sim 7$\,Myr,  the evolutionary history of the WR in the $\gamma^2$\,Vel system seems to have been crafted by some form of binary interactions involving significant mass-loss from the system.
In this context, we note that in the mass-luminosity diagram the location of the WR star is close to the mass-luminosity relation for WR stars as calculated by \citet{SchaererMaeder1992} adopting 
doubled wind mass-loss rates during the main sequence.

\section{Conclusions\label{Conclusion}}

With four new astrometric observations using GRAVITY/VLTI and combined with previous observations with AMBER/VLTI we have obtained a new astrometric orbital solution for the $\gamma^2$\,Vel system. All orbital parameters now have improved  uncertainties by a factor of two to three compared to the previous solution of \citet{Lamberts_etal2017}. 

An independent determination of the orbital period from published spectroscopic orbital velocities allowed solving a remaining ambiguity in the astrometric solution due to the systematic dependence of the parameters $a$, $e$, $\omega$, and $T_0$  on the period. The uncertainty in our improved orbital period $P=78.524\pm 0.002$\,d is smaller by a factor 5 compared to \citet{Schmutz_etal1997}, who determined $P=78.53\pm 0.01$\,d. With the fixed period the co-dependencies of the parameters of the astrometric solution is broken.

From the observed semi-major axis and the measured orbital velocity, we have calculated the distance of the system. The result agrees with the distance to the $\gamma$\,Vel cluster and we conclude that $\gamma^2$\,Vel is a cluster member. The astrometric distance has a larger uncertainty than that of the cluster and we have thus adopted the cluster distance from \citet{MaizApellaniz_etal2022} as the best distance estimate. 

The interferometric observations not only yield the relative astrometric positions but also yields a measurement of the relative brightness of the two stellar components. We find that the WR star is 
${\cal F}_K(\mathrm{WR})/{\cal F}_K(\mathrm{O})=1.46\pm 0.09$ brighter in the K-band than the O star. This is to some extent surprising because the predicted WR to O-star light ratio in the K-band is  that the two components should be about equally bright. The implication of this finding is that most likely, the atmosphere model for the WR star is not correct in the infrared. Trusting more the atmosphere model for the O star, the observations allow to derive the angular diameter of the O star and using the cluster distance we get the radius of the O star 
which, in turn, leads to its luminosity $L_\mathrm{O} = (2.5\pm 0.4)\times 10^5\,\mathrm{L}_\odot$ using the published effective temperature.

\begin{acknowledgements}
We thank Luc Dessart for sending us the spectrum of the WC-star atmosphere model presented in \citet{Milour_etal2007} and \citet{Lamberts_etal2017}.
This research has made use of the SVO Filter Profile Service "Carlos Rodrigo", funded by MCIN/AEI/10.13039/501100011033/ through grant PID2023-146210NB-I00 and of the VizieR catalogue access tool, CDS,
 Strasbourg, France \citep{vizier2000}.
This research has made use of the Jean-Marie Mariotti Center JSDC catalogue, which involves the JSDC catalogue.\footnote{JSDC available at http://www.jmmc.fr/jsdc.}
This work has made use of data from the European Space Agency (ESA) mission Gaia (https://www.cosmos.esa.int/gaia), processed by the Gaia Data Processing and Analysis Consortium (DPAC, https://www.cosmos.esa.int/web/gaia/dpac/consortium). Funding for the DPAC has been provided by national institutions, in particular the institutions participating in the Gaia Multilateral Agreement. GK acknowledges support from UNAM/DGAPA/PAPIIT grant IN105626.
J.S.-B. acknowledges the support received by the UNAM DGAPA-PAPIIT project AG 101025 and from the SECIHTI Ciencia de Frontera project CBF-2025-I-3033.
\end{acknowledgements}

\bibliographystyle{aa} 
\bibliography{aa60042-26_references} 

@ARTICLE{Baade_etal1990,
       author = {{Baade}, D. and {Schmutz}, W. and {van Kerkwijk}, M.},
        title = "{Short-term activity in the gamma2 Velorum system : the O-type supergiant is a nonradially pulsating star.}",
      journal = {\aap},
         year = 1990,
        month = dec,
       volume = {240},
        pages = {105}
}

@ARTICLE{Baroch_etal2021,
       author = {{Baroch}, D. and {Gim{\'e}nez}, A. and {Ribas}, I. and {Morales}, J.~C. and {Anglada-Escud{\'e}}, G. and {Claret}, A.},
        title = "{Analysis of apsidal motion in eclipsing binaries using TESS data. I. A test of gravitational theories}",
      journal = {\aap},
         year = 2021,
        month = may,
       volume = {649},
          eid = {A64},
        pages = {A64},
          doi = {10.1051/0004-6361/202040004},
archivePrefix = {arXiv},
       eprint = {2103.03140},
 primaryClass = {astro-ph.SR},
       adsurl = {https://ui.adsabs.harvard.edu/abs/2021A&A...649A..64B}
}

@ARTICLE{2024MNRAS.528.2026C,
       author = {{Crowther}, Paul A. and {Barlow}, M.~J. and {Royer}, P. and {Hillier}, D.~J. and {Bestenlehner}, J.~M. and {Morris}, P.~W. and {Wesson}, R.},
        title = "{Oxygen abundance of {\ensuremath{\gamma}} Vel from [O III] 88 {\ensuremath{\mu}}m Herschel/PACS spectroscopy}",
      journal = {\mnras},
         year = 2024,
        month = feb,
       volume = {528},
       number = {2},
        pages = {2026-2039},
          doi = {10.1093/mnras/stae145},
archivePrefix = {arXiv},
       eprint = {2310.15170},
 primaryClass = {astro-ph.SR},
       adsurl = {https://ui.adsabs.harvard.edu/abs/2024MNRAS.528.2026C}
}

@ARTICLE{Davis_etal1999,
       author = {{Davis}, J. and {Tango}, W.~J. and {Booth}, A.~J. and {ten Brummelaar}, T.~A. and {Minard}, R.~A. and {Owens}, S.~M.},
        title = "{The Sydney University Stellar Interferometer - I. The instrument}",
      journal = {\mnras},
         year = 1999,
        month = mar,
       volume = {303},
       number = {4},
        pages = {773-782},
          doi = {10.1046/j.1365-8711.1999.02269.x}
}

@ARTICLE{deMarco_etalI1999,
       author = {{De Marco}, Orsola and {Schmutz}, W.},
        title = "{The {\ensuremath{\gamma}} Velorum binary system. I. O star parameters and light ratio}",
      journal = {\aap},
         year = 1999,
        month = may,
       volume = {345},
        pages = {163-171}
}

@INPROCEEDINGS{DeMarco2002,
       author = {{De Marco}, Orsola},
        title = "{Optical Line-Variability in the O+WR Binary {\ensuremath{\gamma}} Vel}",
    booktitle = {Interacting Winds from Massive Stars},
         year = 2002,
       editor = {{Moffat}, Anthony F.~J. and {St-Louis}, Nicole},
       series = {Astronomical Society of the Pacific Conference Series},
       volume = {260},
        month = jan,
        pages = {517}
}

@ARTICLE{deMarco_etalII2000,
       author = {{De Marco}, O. and {Schmutz}, W. and {Crowther}, P.~A. and {Hillier}, D.~J. and {Dessart}, L. and {de Koter}, A. and {Schweickhardt}, J.},
        title = "{The {\ensuremath{\gamma}} Velorum binary system. II. WR stellar parameters and the photon loss mechanism}",
      journal = {\aap},
         year = 2000,
        month = jun,
       volume = {358},
        pages = {187-200},
          doi = {10.48550/arXiv.astro-ph/0004081}
}

@Misc{Dessart2025,
  author       = {{Dessart}, Luc},
  howpublished = {{Personal Communication}},
  year         = {2025},
}

@ARTICLE{Dessart_etal2000,
       author = {{Dessart}, Luc and {Crowther}, Paul A. and {Hillier}, D. John and {Willis}, Allan J. and {Morris}, Patrick W. and {van der Hucht}, Karel A.},
        title = "{Quantitative analysis of WC stars: constraints on neon abundances from ISO-SWS spectroscopy}",
      journal = {\mnras},
         year = 2000,
        month = jun,
       volume = {315},
       number = {2},
        pages = {407-422},
          doi = {10.1046/j.1365-8711.2000.03399.x}
}

@ARTICLE{Ekstrom_etal2012,
       author = {{Ekstr{\"o}m}, S. and {Georgy}, C. and {Eggenberger}, P. and
         {Meynet}, G. and {Mowlavi}, N. and {Wyttenbach}, A. and {Granada}, A. and
         {Decressin}, T. and {Hirschi}, R. and {Frischknecht}, U. and
         {Charbonnel}, C. and {Maeder}, A.},
        title = "{Grids of stellar models with rotation. I. Models from 0.8 to 120 M$_{☉}$ at solar metallicity (Z = 0.014)}",
      journal = {\aap},
         year = "2012",
        month = "Jan",
       volume = {537},
          eid = {A146},
        pages = {A146},
          doi = {10.1051/0004-6361/201117751},
}

@ARTICLE{2009MNRAS.400L..20E,
       author = {{Eldridge}, John J.},
        title = "{A new-age determination for {\ensuremath{\gamma}}$^{2}$ Velorum from binary stellar evolution models}",
      journal = {\mnras},
         year = 2009,
        month = nov,
       volume = {400},
       number = {1},
        pages = {L20-L23},
          doi = {10.1111/j.1745-3933.2009.00753.x},
archivePrefix = {arXiv},
       eprint = {0909.0504},
 primaryClass = {astro-ph.SR},
       adsurl = {https://ui.adsabs.harvard.edu/abs/2009MNRAS.400L..20E}
}

@PROCEEDINGS{ESA1997,
        title = "{The HIPPARCOS and TYCHO catalogues. Astrometric and photometric star catalogues derived from the ESA HIPPARCOS Space Astrometry Mission}",
       author = {{ESA},},
    booktitle = {ESA Special Publication},
         year = 1997,
       series = {ESA Special Publication},
           editor = {ESA},
       volume = {1200},
        month = jan,
       adsurl = {https://ui.adsabs.harvard.edu/abs/1997ESASP1200.....E},
}

@ARTICLE{2004A&A...423..267G,
       author = {{Georgiev}, L.~N. and {Koenigsberger}, G.},
        title = "{Line profile variations in WR+O binary systems. I. The code and basic predictions}",
      journal = {\aap},
         year = 2004,
        month = aug,
       volume = {423},
        pages = {267-279},
          doi = {10.1051/0004-6361:200400030},
       adsurl = {https://ui.adsabs.harvard.edu/abs/2004A&A...423..267G}
}

@ARTICLE{1970MNRAS.148..103H,
       author = {{Hanbury Brown}, R. and {Davis}, J. and {Herbison-Evans}, D. and {Allen}, L.~R.},
        title = "{A study of {\ensuremath{\gamma}}$^{2}$ Velorum with a stellar intensity interferometer.}",
      journal = {\mnras},
         year = 1970,
        month = jan,
       volume = {148},
        pages = {103-117},
          doi = {10.1093/mnras/148.1.103},
       adsurl = {https://ui.adsabs.harvard.edu/abs/1970MNRAS.148..103H}
}

@ARTICLE{Henley_etal2005,
       author = {{Henley}, David B. and {Stevens}, Ian R. and {Pittard}, Julian M.},
        title = "{Probing the wind-wind collision in {\ensuremath{\gamma}}$^{2}$ Velorum with high-resolution Chandra X-ray spectroscopy: evidence for sudden radiative braking and non-equilibrium ionization}",
      journal = {\mnras},
         year = 2005,
        month = feb,
       volume = {356},
       number = {4},
        pages = {1308-1326},
          doi = {10.1111/j.1365-2966.2004.08556.x}
}

@PROCEEDINGS{Herrero_etal1992,
       author = {{Herrero}, A. and {Kudritzki}, R.~P. and {Vilchez}, J.~M. and {Kunze}, D. and {Butler}, K. and {Haser}, S.},
        title = "{The mass and helium discrepancy in massive young stars}",
    booktitle = {The Atmospheres of Early-Type Stars},
         year = 1992,
       editor = {{Heber}, Ulrich and {Jeffery}, C. Simon},
       volume = {401},
        pages = {21},
          doi = {10.1007/3-540-55256-1_269}
}

@ARTICLE{2000MNRAS.318..402H,
       author = {{Hill}, G.~M. and {Moffat}, A.~F.~J. and {St-Louis}, N. and {Bartzakos}, P.},
        title = "{Modelling the spectra of colliding winds in the Wolf-Rayet WC7+O binaries WR 42 and WR 79}",
      journal = {\mnras},
         year = 2000,
        month = oct,
       volume = {318},
       number = {2},
        pages = {402-410},
          doi = {10.1046/j.1365-8711.2000.03705.x},
       adsurl = {https://ui.adsabs.harvard.edu/abs/2000MNRAS.318..402H}
}

@ARTICLE{Holdsworth_etal2024,
       author = {{Holdsworth}, Amanda and {Richardson}, Noel and {Schaefer}, Gail H. and {Eldridge}, Jan J. and {Hill}, Grant M. and {Spejcher}, Becca and {Mackey}, Jonathan and {Moffat}, Anthony F.~J. and {Navarete}, Felipe and {Monnier}, John D. and {Kraus}, Stefan and {Le Bouquin}, Jean-Baptiste and {Anugu}, Narsireddy and {Chhabra}, Sorabh and {Codron}, Isabelle and {Ennis}, Jacob and {Gardner}, Tyler and {Gutierrez}, Mayra and {Ibrahim}, Noura and {Labdon}, Aaron and {Lanthermann}, Cyprien and {Setterholm}, Benjamin R.},
        title = "{Visual Orbits of Wolf─Rayet Stars. II. The Orbit of the Nitrogen-rich Wolf─Rayet Binary WR 138 Measured with the CHARA Array}",
      journal = {\apj},
         year = 2024,
        month = dec,
       volume = {977},
       number = {2},
          eid = {185},
        pages = {185},
          doi = {10.3847/1538-4357/ad9024}
}

@ARTICLE{2009MNRAS.395..962I,
       author = {{Ignace}, R. and {Bessey}, R. and {Price}, C.~S.},
        title = "{Modelling forbidden line emission profiles from colliding wind binaries}",
      journal = {\mnras},
         year = 2009,
        month = may,
       volume = {395},
       number = {2},
        pages = {962-972},
          doi = {10.1111/j.1365-2966.2009.14586.x},
archivePrefix = {arXiv},
       eprint = {0902.0527},
 primaryClass = {astro-ph.SR},
       adsurl = {https://ui.adsabs.harvard.edu/abs/2009MNRAS.395..962I}
}

@ARTICLE{Kimeswenger_etal2004,
       author = {{Kimeswenger}, S. and {Lederle}, C. and {Richichi}, A. and {Percheron}, I. and {Paresce}, F. and {Armsdorfer}, B. and {Bacher}, A. and {Cabrera-Lavers}, A.~L. and {Kausch}, W. and {Rassia}, E. and {Schmeja}, S. and {Tapken}, C. and {Fouqu{\'e}}, P. and {Maury}, A. and {Epchtein}, N.},
        title = "{J - K DENIS photometry of a VLTI-selected sample of bright southern stars}",
      journal = {\aap},
         year = 2004,
        month = jan,
       volume = {413},
        pages = {1037-1043},
          doi = {10.1051/0004-6361:20031576}
}

@ARTICLE{Kurosawa_etal2002,
       author = {{Kurosawa}, R. and {Hillier}, D.~J. and {Pittard}, J.~M.},
        title = "{Mass-loss rate determination for the massive binary V444 Cygni using 3-D Monte-Carlo simulations of line and polarization variability}",
      journal = {\aap},
         year = 2002,
        month = jun,
       volume = {388},
        pages = {957-977},
          doi = {10.1051/0004-6361:20020443}
}

@ARTICLE{Lamberts_etal2017,
   author = {{Lamberts}, A. and {Millour}, F. and {Liermann}, A. and {Dessart}, L. and 
	{Driebe}, T. and {Duvert}, G. and {Finsterle}, W. and {Girault}, V. and 
	{Massi}, F. and {Petrov}, R.~G. and {Schmutz}, W. and {Weigelt}, G. and 
	{Chesneau}, O.},
    title = "{Numerical simulations and infrared spectro-interferometry reveal the wind collision region in {$\gamma$}$^{2}$ Velorum}",
  journal = {\mnras},
archivePrefix = "arXiv",
   eprint = {1701.01124},
 primaryClass = "astro-ph.SR",
     year = 2017,
    month = jul,
   volume = 468,
    pages = {2655-2671},
      doi = {10.1093/mnras/stx588}
}

@ARTICLE{Luehrs_1997,
       author = {{L\"uhrs}, S.},
        title = "{A Colliding-Wind Model for the Wolf-Rayet System HD 152270}",
      journal = {\pasp},
         year = 1997,
        month = may,
       volume = {109},
        pages = {504-513},
          doi = {10.1086/133907},
       adsurl = {https://ui.adsabs.harvard.edu/abs/1997PASP..109..504L}
}

@ARTICLE{2012ApJ...748...96M,
       author = {{Massey}, Philip and {Morrell}, Nidia I. and {Neugent}, Kathryn F. and {Penny}, Laura R. and {DeGioia-Eastwood}, Kathleen and {Gies}, Douglas R.},
        title = "{Photometric and Spectroscopic Studies of Massive Binaries in the Large Magellanic Cloud. I. Introduction and Orbits for Two Detached Systems: Evidence for a Mass Discrepancy?}",
      journal = {\apj},
         year = 2012,
        month = apr,
       volume = {748},
       number = {2},
          eid = {96},
        pages = {96},
          doi = {10.1088/0004-637X/748/2/96},
archivePrefix = {arXiv},
       eprint = {1201.3280},
 primaryClass = {astro-ph.SR},
       adsurl = {https://ui.adsabs.harvard.edu/abs/2012ApJ...748...96M}
}

@ARTICLE{Milour_etal2007,
       author = {{Millour}, F. and {Petrov}, R.~G. and {Chesneau}, O. and {Bonneau}, D. and {Dessart}, L. and {Bechet}, C. and {Tallon-Bosc}, I. and {Tallon}, M. and {Thi{\'e}baut}, E. and {Vakili}, F. and {Malbet}, F. and {Mourard}, D. and {Antonelli}, P. and {Beckmann}, U. and {Bresson}, Y. and {Chelli}, A. and {Dugu{\'e}}, M. and {Duvert}, G. and {Gennari}, S. and {Gl{\"u}ck}, L. and {Kern}, P. and {Lagarde}, S. and {Le Coarer}, E. and {Lisi}, F. and {Perraut}, K. and {Puget}, P. and {Rantakyr{\"o}}, F. and {Robbe-Dubois}, S. and {Roussel}, A. and {Tatulli}, E. and {Weigelt}, G. and {Zins}, G. and {Accardo}, M. and {Acke}, B. and {Agabi}, K. and {Altariba}, E. and {Arezki}, B. and {Aristidi}, E. and {Baffa}, C. and {Behrend}, J. and {Bl{\"o}cker}, T. and {Bonhomme}, S. and {Busoni}, S. and {Cassaing}, F. and {Clausse}, J. -M. and {Colin}, J. and {Connot}, C. and {Delboulb{\'e}}, A. and {Domiciano de Souza}, A. and {Driebe}, T. and {Feautrier}, P. and {Ferruzzi}, D. and {Forveille}, T. and {Fossat}, E. and {Foy}, R. and {Fraix-Burnet}, D. and {Gallardo}, A. and {Giani}, E. and {Gil}, C. and {Glentzlin}, A. and {Heiden}, M. and {Heininger}, M. and {Hernandez Utrera}, O. and {Hofmann}, K. -H. and {Kamm}, D. and {Kiekebusch}, M. and {Kraus}, S. and {Le Contel}, D. and {Le Contel}, J. -M. and {Lesourd}, T. and {Lopez}, B. and {Lopez}, M. and {Magnard}, Y. and {Marconi}, A. and {Mars}, G. and {Martinot-Lagarde}, G. and {Mathias}, P. and {M{\`e}ge}, P. and {Monin}, J. -L. and {Mouillet}, D. and {Nussbaum}, E. and {Ohnaka}, K. and {Pacheco}, J. and {Perrier}, C. and {Rabbia}, Y. and {Rebattu}, S. and {Reynaud}, F. and {Richichi}, A. and {Robini}, A. and {Sacchettini}, M. and {Schertl}, D. and {Sch{\"o}ller}, M. and {Solscheid}, W. and {Spang}, A. and {Stee}, P. and {Stefanini}, P. and {Tasso}, D. and {Testi}, L. and {von der L{\"u}he}, O. and {Valtier}, J. -C. and {Vannier}, M. and {Ventura}, N.},
        title = "{Direct constraint on the distance of {\ensuremath{\gamma}}$^{2}$ Velorum from AMBER/VLTI observations}",
      journal = {\aap},
         year = 2007,
        month = mar,
       volume = {464},
       number = {1},
        pages = {107-118},
          doi = {10.1051/0004-6361:20065408}
}

@ARTICLE{Munch1950,
       author = {{M{\"u}nch}, Guido},
        title = "{A Spectrographic Study of HD 193576.}",
      journal = {\apj},
         year = 1950,
        month = sep,
       volume = {112},
        pages = {266},
          doi = {10.1086/145341}
}

@ARTICLE{Niemela_etal1980,
       author = {{Niemela}, V.~S. and {Sahade}, J.},
        title = "{The orbital elements of gam2 Vel.}",
      journal = {\apj},
         year = 1980,
        month = may,
       volume = {238},
        pages = {244-249},
          doi = {10.1086/157981}
}

@ARTICLE{North_etal2007,
       author = {{North}, J.~R. and {Tuthill}, P.~G. and {Tango}, W.~J. and {Davis}, J.},
        title = "{{\ensuremath{\gamma}}$^{2}$ Velorum: orbital solution and fundamental parameter determination with SUSI}",
      journal = {\mnras},
         year = 2007,
        month = may,
       volume = {377},
       number = {1},
        pages = {415-424},
          doi = {10.1111/j.1365-2966.2007.11608.x}
}

@ARTICLE{vizier2000,
       author = {{Ochsenbein}, F. and {Bauer}, P. and {Marcout}, J.},
        title = "{The VizieR database of astronomical catalogues}",
      journal = {\aaps},
         year = 2000,
        month = apr,
       volume = {143},
        pages = {23-32},
          doi = {10.1051/aas:2000169}
          }

@ARTICLE{Perrine1920,
       author = {{Perrine}, C.~D.},
        title = "{Temporary shifting absorption at the heads of helium bands in the spectrum of {\ensuremath{\gamma}} Argus.}",
      journal = {\apj},
         year = 1920,
        month = jul,
       volume = {52},
        pages = {39-46},
          doi = {10.1086/142557}
}

@ARTICLE{Pike_etal1983,
       author = {{Pike}, C.~D. and {Stickland}, D.~J. and {Willis}, A.~J.},
        title = "{The orbit of gamma 2 Velorum}",
      journal = {The Observatory},
         year = 1983,
        month = jun,
       volume = {103},
        pages = {154-159}
}

@ARTICLE{2021ApJ...908L...3R,
       author = {{Richardson}, Noel D. and {Lee}, Laura and {Schaefer}, Gail and {Shenar}, Tomer and {Sander}, Andreas A.~C. and {Hill}, Grant M. and {Fullard}, Andrew G. and {Monnier}, John D. and {Anugu}, Narsireddy and {Davies}, Claire L. and et al.},
        title = "{The First Dynamical Mass Determination of a Nitrogen-rich Wolf-Rayet Star Using a Combined Visual and Spectroscopic Orbit}",
      journal = {\apjl},
         year = 2021,
        month = feb,
       volume = {908},
       number = {1},
          eid = {L3},
        pages = {L3},
          doi = {10.3847/2041-8213/abd722},
archivePrefix = {arXiv},
       eprint = {2101.04232},
 primaryClass = {astro-ph.SR},
       adsurl = {https://ui.adsabs.harvard.edu/abs/2021ApJ...908L...3R}
}

@ARTICLE{Richardson_etal2017,
   author = {{Richardson}, N.~D. and {Russell}, C.~M.~P. and {St-Jean}, L. and 
	{Moffat}, A.~F.~J. and {St-Louis}, N. and {Shenar}, T. and {Pablo}, H. and 
	{Hill}, G.~M. and {Ramiaramanantsoa}, T. and {Corcoran}, M. and 
	{Hamuguchi}, K. and {Eversberg}, T. and {Miszalski}, B. and 
	{Chen{\'e}}, A.-N. and {Waldron}, W. and {Kotze}, E.~J. and 
	{Kotze}, M.~M. and {Luckas}, P. and {Cacella}, P. and {Heathcote}, B. and 
	{Powles}, J. and {Bohlsen}, T. and {Locke}, M. and {Handler}, G. and 
	{Kuschnig}, R. and {Pigulski}, A. and {Popowicz}, A. and {Wade}, G.~A. and 
	{Weiss}, W.~W.},
    title = "{The variability of the BRITE-est Wolf-Rayet binary, {$\gamma$}$^{2}$ Velorum-I. Photometric and spectroscopic evidence for colliding winds}",
  journal = {\mnras},
archivePrefix = "arXiv",
   eprint = {1707.03390},
 primaryClass = "astro-ph.SR",
     year = 2017,
    month = nov,
   volume = 471,
    pages = {2715-2729},
      doi = {10.1093/mnras/stx1731}
}

@ARTICLE{Richardson_etal2024,
       author = {{Richardson}, Noel D. and {Schaefer}, Gail H. and {Eldridge}, Jan J. and {Spejcher}, Rebecca and {Holdsworth}, Amanda and {Lau}, Ryan M. and {Monnier}, John D. and {Moffat}, Anthony F.~J. and {Weigelt}, Gerd and {Williams}, Peredur M. and {Kraus}, Stefan and {Le Bouquin}, Jean-Baptiste and {Anugu}, Narsireddy and {Chhabra}, Sorabh and {Codron}, Isabelle and {Ennis}, Jacob and {Gardner}, Tyler and {Gutierrez}, Mayra and {Ibrahim}, Noura and {Labdon}, Aaron and {Lanthermann}, Cyprien and {Setterholm}, Benjamin R.},
        title = "{Visual Orbits of Wolf─Rayet Stars. I. The Orbit of the Dust-producing Wolf─Rayet Binary WR 137 Measured with the CHARA Array}",
      journal = {\apj},
         year = 2024,
        month = dec,
       volume = {977},
       number = {1},
          eid = {78},
        pages = {78},
          doi = {10.3847/1538-4357/ad8d5c}
}

@ARTICLE{Richardson_etal2016,
       author = {{Richardson}, Noel D. and {Shenar}, Tomer and {Roy-Loubier}, Olivier and {Schaefer}, Gail and {Moffat}, Anthony F.~J. and {St-Louis}, Nicole and {Gies}, Douglas R. and {Farrington}, Chris and {Hill}, Grant M. and {Williams}, Peredur M. and {Gordon}, Kathryn and {Pablo}, Herbert and {Ramiaramanantsoa}, Tahina},
        title = "{The CHARA Array resolves the long-period Wolf-Rayet binaries WR 137 and WR 138}",
      journal = {\mnras},
         year = 2016,
        month = oct,
       volume = {461},
       number = {4},
        pages = {4115-4124},
          doi = {10.1093/mnras/stw1585}
}

@ARTICLE{Sander_etal2012,
       author = {{Sander}, A. and {Hamann}, W.-R. and {Todt}, H.},
        title = "{The Galactic WC stars. Stellar parameters from spectral analyses indicate a new evolutionary sequence}",
      journal = {\aap},
         year = 2012,
        month = apr,
       volume = {540},
          eid = {A144},
        pages = {A144},
          doi = {10.1051/0004-6361/201117830}
}

@ARTICLE{SchaererMaeder1992,
       author = {{Schaerer}, D. and {Maeder}, A.},
        title = "{Basic relations between physical parameters of Wolf-Rayet stars}",
      journal = {\aap},
         year = 1992,
        month = sep,
       volume = {263},
       number = {1-2},
        pages = {129-136}
}

@ARTICLE{Schild_etal2004,
       author = {{Schild}, H. and {G{\"u}del}, M. and {Mewe}, R. and {Schmutz}, W. and {Raassen}, A.~J.~J. and {Audard}, M. and {Dumm}, T. and {van der Hucht}, K.~A. and {Leutenegger}, M.~A. and {Skinner}, S.~L.},
        title = "{Wind clumping and the wind-wind collision zone  in the Wolf-Rayet binary {\ensuremath{\gamma}}$^{2}$ Velorum observations at high and low state. XMM-Newton observations at high and low state}",
      journal = {\aap},
         year = 2004,
        month = jul,
       volume = {422},
        pages = {177-191},
          doi = {10.1051/0004-6361:20047035}
}

@article{Schmutz1997,
    author = {{Schmutz}, W.},
    title = "{Photon loss from the helium Ly{$\alpha$} line - the key to the acceleration of Wolf-Rayet winds.}",
  journal = {\aap},
     year = 1997,
    month = may,
   volume = 321,
    pages = {268-287}
}

@ARTICLE{Schmutz_etal1997,
       author = {{Schmutz}, W. and {Schweickhardt}, J. and {Stahl}, O. and {Wolf}, B. and {Dumm}, T. and {Gang}, Th. and {Jankovics}, I. and {Kaufer}, A. and {Lehmann}, H. and {Mandel}, H. and {Peitz}, J. and {Rivinius}, Th.},
        title = "{The orbital motion of gamma\^2 Velorum}",
      journal = {\aap},
         year = 1997,
        month = dec,
       volume = {328},
        pages = {219-228}
}

@ARTICLE{Skinner_etal2001,
       author = {{Skinner}, Stephen L. and {G{\"u}del}, Manuel and {Schmutz}, Werner and {Stevens}, Ian R.},
        title = "{Chandra Detection of a Close X-Ray Companion and Rich Emission-Line Spectrum in the Wolf-Rayet Binary {\ensuremath{\gamma}} Velorum}",
      journal = {\apjl},
         year = 2001,
        month = sep,
       volume = {558},
       number = {2},
        pages = {L113-L116},
          doi = {10.1086/323567}
}

@ARTICLE{Stevens_etal1992,
       author = {{Stevens}, Ian R. and {Blondin}, John M. and {Pollock}, A.~M.~T.},
        title = "{Colliding Winds from Early-Type Stars in Binary Systems}",
      journal = {\apj},
         year = 1992,
        month = feb,
       volume = {386},
        pages = {265},
          doi = {10.1086/171013}
}

@ARTICLE{Stevens_etal1996,
       author = {{Stevens}, I.~R. and {Corcoran}, M.~F. and {Willis}, A.~J. and {Skinner}, S.~L. and {Pollock}, A.~M.~T. and {Nagase}, F. and {Koyama}, K.},
        title = "{ASCA observations of {\ensuremath{\gamma}}$^{2}$Velorum (WC8+O9I): the variable X-ray spectrum of colliding winds.}",
      journal = {\mnras},
         year = 1996,
        month = dec,
       volume = {283},
       number = {2},
        pages = {589-605},
          doi = {10.1093/mnras/283.2.589}
}

@ARTICLE{2021MNRAS.503..643W,
       author = {{Williams}, Peredur M. and {Varricatt}, Watson P. and {Chen{\'e}}, Andr{\'e}-Nicolas and {Corcoran}, Michael F. and {Gull}, Ted R. and {Hamaguchi}, Kenji and {Moffat}, Anthony F.~J. and {Pollock}, Andrew M.~T. and {Richardson}, Noel D. and {Russell}, Christopher M.~P. and {Sander}, Andreas A.~C. and {Stevens}, Ian R. and {Weigelt}, Gerd},
        title = "{Conditions in the WR 140 wind-collision region revealed by the 1.083-{\ensuremath{\mu}} m He I line profile}",
      journal = {\mnras},
         year = 2021,
        month = may,
       volume = {503},
       number = {1},
        pages = {643-659},
          doi = {10.1093/mnras/stab508},
archivePrefix = {arXiv},
       eprint = {2102.09445},
 primaryClass = {astro-ph.SR},
       adsurl = {https://ui.adsabs.harvard.edu/abs/2021MNRAS.503..643W}
}

@ARTICLE{MaizApellaniz_etal2022,
       author = {{Ma{\'\i}z Apell{\'a}niz}, J. and {Barb{\'a}}, R.~H. and {Fern{\'a}ndez Aranda}, R. and {Pantaleoni Gonz{\'a}lez}, M. and {Crespo Bellido}, P. and {Sota}, A. and {Alfaro}, E.~J.},
        title = "{The Villafranca catalog of Galactic OB groups. II. From Gaia DR2 to EDR3 and ten new systems with O stars}",
      journal = {\aap},
         year = 2022,
        month = jan,
       volume = {657},
          eid = {A131},
        pages = {A131},
          doi = {10.1051/0004-6361/202142364}
}

@ARTICLE{Willis_etal1995,
       author = {{Willis}, A.~J. and {Schild}, H. and {Stevens}, I.~R.},
        title = "{ROSAT observations of {\ensuremath{\gamma}} Velorum (WC8+O9I). I. The discovery of colliding-wind X-ray emission.}",
      journal = {\aap},
         year = 1995,
        month = jun,
       volume = {298},
        pages = {549}
}

\begin{appendix} 
\nolinenumbers 

\section{Enlargement of the differences between VLTI measurements and fitted orbit}
\begin{figure*}[t!]
   \centering
   \includegraphics[width=18cm]{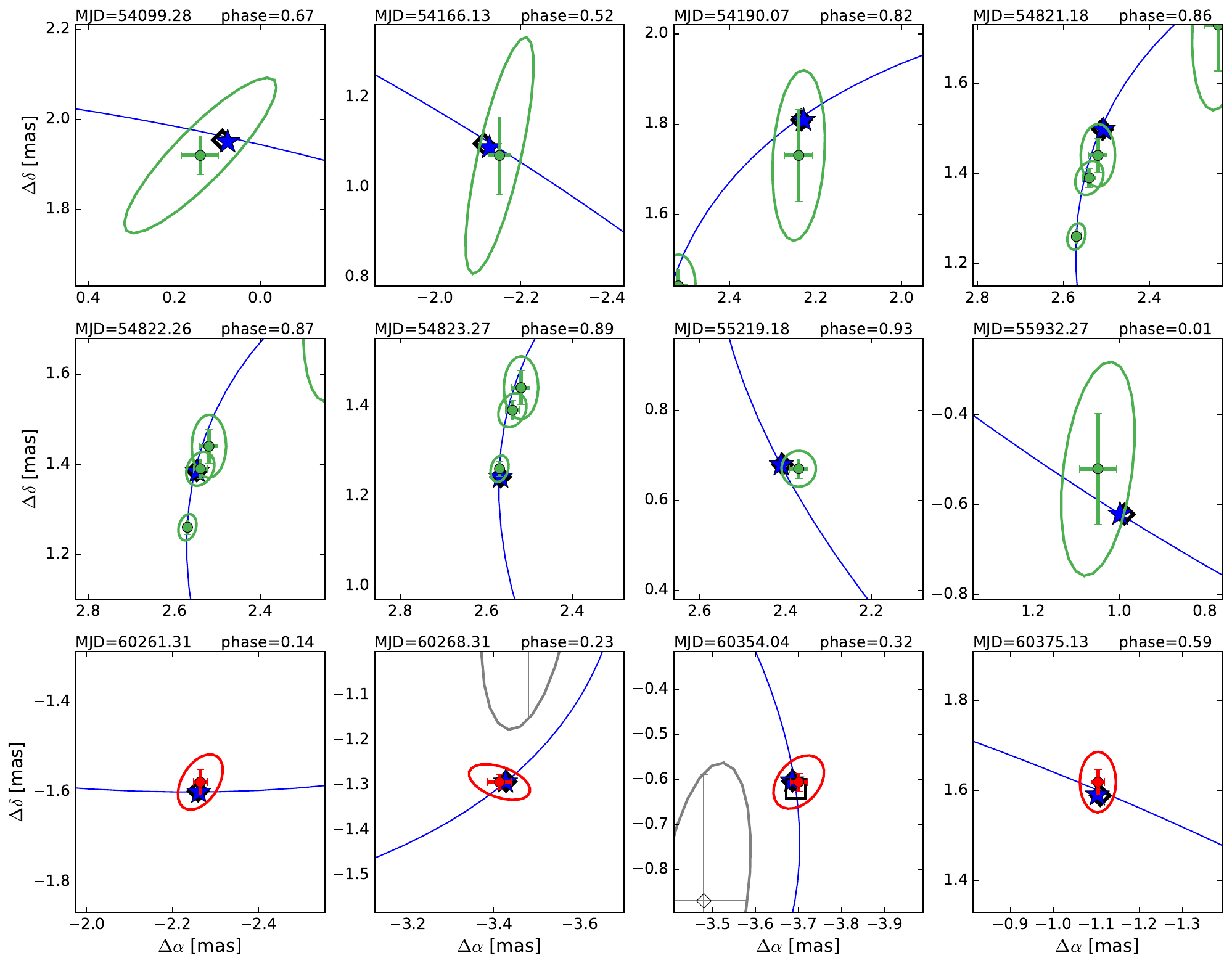}
   \centering
   \caption{As Fig.\,1 but enlargements around the positions of the measurements. The green error bars mark the 1\,$\sigma$ uncertainties of the AMBER observations as estimated in this paper whereas the green ellipses indicate the uncertainties as given by \citet{Lamberts_etal2017} (see text Sect.\, 3.1). The red ellipses mark the 2.5\,$\sigma$ range as given in Tab.\,\ref{tab:GRAVITY_observations} and the red error bars are of 1\,$\sigma$ length. The blue stars are the calculated positions of the best orbital solution. The open magenta diamonds denote the positions calculated with a fixed period of $P=78.524$\,d. The gray ellipse in the panels with phase 0.23 and 0.32 gives the uncertainty limit of the position measured by AMBER/VLTI in December 2004. The open grey square in the panel with phase 0.32 marks the calculated position for MJD=53365.19, i.e. 25 December 2004.}
              \label{Fig:A1}%
    \end{figure*}

In Fig.\,A1 we illustrate enlargements of Fig.\,1 around the measurements, such that the differences between observations and orbital solution are apparent.

\section{Determining the orbital period}
The orbital period of $\gamma^2$\,Vel is calculated using two methods. 

The first one is fitting an orbital velocity solution to the reported radial velocities with fixed orbital parameters for the eccentricity, $e=0.322$, the periastron longitude, $\omega=67.6^\circ$, and period, $P=78.524\,$d and solve for the velocity amplitude, $K$, the system velocity $v_0$, and the periastron passage time, $T_0$. As $T_0$ is referred to a given $\omega$ it can be compared to the $T_0$ determined from the astrometry reported in this paper. 
The uncertainty of a period determination is given by the sum of the precisions of the two $T_0$ epochs divided by the number of orbits between the two epochs.
Table B.1 lists five periods determined with method\,1. The standard deviation of the five period values is $\sigma_m= 0.0023$\,d. As this standard deviation is 31\,\%  larger than the mean uncertainties determined for the individual periods, we increase the listed precisions by this amount.
The weighted mean is $78.5246\pm 0.0009$\,d, with the given uncertainty the standard error of the weighted mean, with squared enlarged inverse-variance weights.

The second method is fitting an orbit solution to the combined data set of measured radial velocities from two epochs. 
As in method\,1 we fix the orbital parameters for the eccentricity, $e=0.322$ and the periastron longitude, $\omega=67.5^\circ$, but in this method the period is a free parameter, which is fitted. The listed uncertainty is derived from the covariance matrix, which is returned by the {\em optimize} routine. 
\citet{Richardson_etal2017} have published radial velocity measurements of numerous emission and absorption lines, whereas older publications give mean radial velocities for emission and absorption lines. There are many different combinations of radial velocity pairs possible. In Tab.\,B.1 we list seven selected combinations. 
Because different lines yield substantially different system velocities we subtract the system velocities before combining the data sets. The standard deviation of the seven periods is $\sigma_m= 0.0030$\,d, which is more than a factor two larger than the precisions listed in Tab.\,B.1, which have a mean error of $\sigma_{\bar{i}} = 0.0013$\,d. Thus, systematic influences dominate the periods calculated with method\,2. Therefore, we assume a constant uncertainty given by the standard deviation and derive a mean period for method\,2 of $78.5231\pm 0.0013$\,d, with the uncertainty given by the standard deviation divided by the square root of six.

\begin{table*}
      \caption[]{Results of period determinations with method\,1 or 2 (see text Appendix B). Radial velocities have been published by \citet{Niemela_etal1980}(N80), \citet{Pike_etal1983} (P83), and \citet{Richardson_etal2017} (R17).}
         \label{Pestimates}
     $$
         \begin{array}{cclclrl}
            \hline
            \noalign{\smallskip}
            \mathrm{Method} &             \mathrm{Epoch\,1\ [MJD]} & \mathrm{Source\,1} & 
            \mathrm{Epoch\,2\ [MJD]} & \mathrm{Source\,2} &
            \mathrm{Orbits} & P\, {[\mathrm{d}]} \\
            \hline
            \noalign{\smallskip}
            1 & 60250.44\pm 0.08 & \mathrm{VLTI}& 
            32846.37\pm 0.46 & \mathrm{N80\ He\,\textsc{ii}}\,\mathrm{em} & 
            349 & 78.5221\pm 0.0018\\
            2 & - & \mathrm{R17\ C\,\textsc{iv}}\,\lambda 5804 & 
            32846.74\pm 0.55   &  \mathrm{N80\ He\,\textsc{ii}}\,\mathrm{em} & 
              & 78.5225\pm 0.0012 \\
            \noalign{\smallskip}
            1 & 60250.44\pm 0.16 & \mathrm{VLTI}& 
            32844.59\pm 0.49 & \mathrm{N80\, abs} &
            349 & 78.5272\pm 0.0019\\
            2 & - & \mathrm{R17\ H\,\beta\,abs}& 
            32844.54\pm 0.28   &  \mathrm{N80\, abs} &  & 
            78.5229\pm 0.0012 \\
            \noalign{\smallskip}
            1 & 60250.44\pm 0.16 & \mathrm{VLTI}& 
            32844.35\pm 0.62 & \mathrm{N80\, C\,em} &
            349 & 78.5279\pm 0.0022\\
            2 & - & \mathrm{R17\ C\,\textsc{iv}}\,\lambda 5804& 
            32845.35\pm 0.54  &  \mathrm{N80\,C\,em} &  & 
            78.5301\pm 0.0012 \\
            2 & - & \mathrm{R17\ C\,\textsc{iii}}\,\lambda 5696& 
            32844.69\pm 0.25   &  \mathrm{N80\,C\,em} &  & 
            78.5210\pm 0.0007  \\
            \noalign{\smallskip}
            1 & 60250.44\pm 0.16 & \mathrm{VLTI}& 
            35201.74\pm 0.23 & \mathrm{P83\,WR\,em} &
            319 & 78.5231\pm 0.0012\\
            2 & - & \mathrm{R17\ C\,\textsc{iv}\,\lambda 5804 }& 
            35201.71\pm 0.47 &  \mathrm{P83\,WR\,em} &  & 
            78.5231\pm 0.0018  \\
            2 & - & \mathrm{R17\ C\,\textsc{iii}\,\lambda 5696 }& 
            35201.61\pm 0.30 &  \mathrm{P83\,WR\,em} &  & 
            78.5216\pm 0.0011  \\
            \noalign{\smallskip}
            1 & 60250.44\pm 0.16 & \mathrm{VLTI}& 
            17611.18\pm 0.67 & \mathrm{P83\,O\,abs} &
            543 & 78.5256\pm 0.0015\\
            2 & - & \mathrm{R17\ H\,\beta\,abs}& 
            17612.66\pm 0.71   &  \mathrm{P83\,O\,abs} &  & 
            78.5208\pm 0.0016 \\
            \hline
         \end{array}
     $$
   \end{table*}

\section{Re-evaluation of the radial velocity measurements}
\begin{figure*}
   \centering
   \includegraphics[width=6cm]{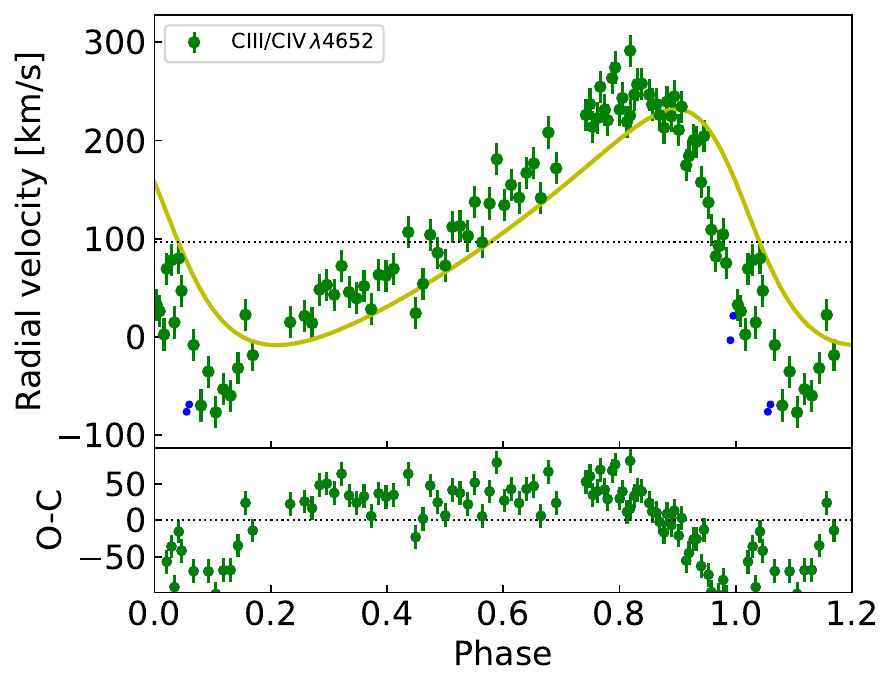}
   \includegraphics[width=6cm]{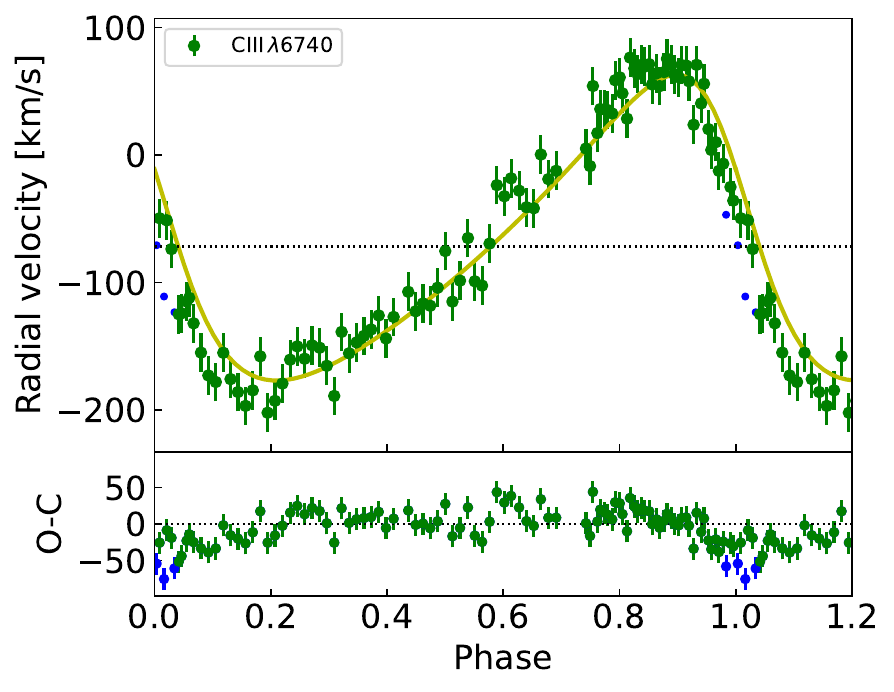}
   \includegraphics[width=6cm]{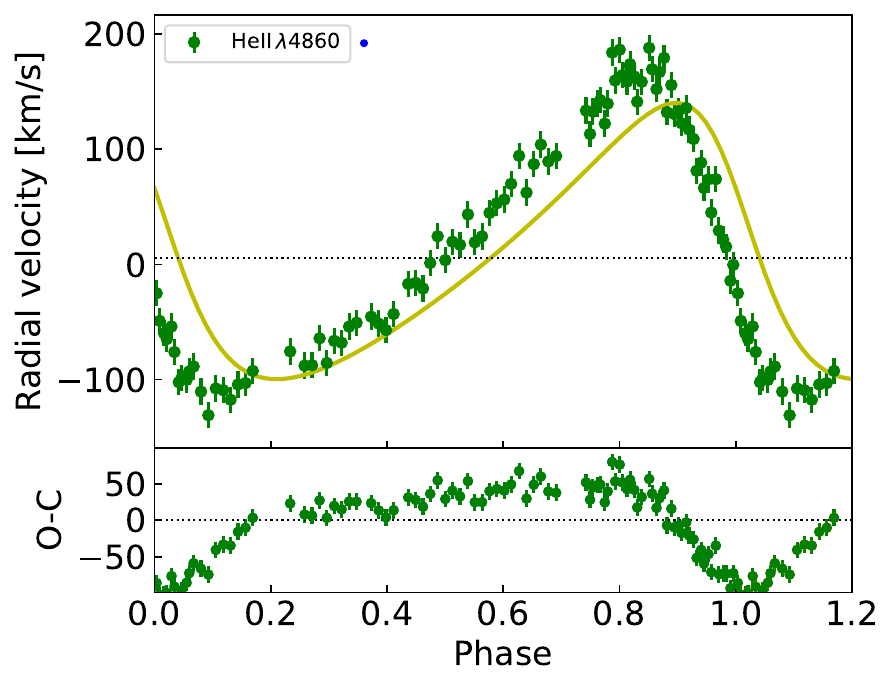}
   \centering
   \caption{Comparison of the radial velocity measurements to the predicted solution using the astrometric parameters from the column with $P=78.524$\,d in Tab.\,\ref{tab:orbit} and given velocity amplitude $K=120$\,km/s. The only free parameter is $V_0$. All three line measurements exhibit a shift to an earlier phase 
   and larger velocity amplitudes.
           }
              \label{Fig:C1}%
    \end{figure*}
\begin{figure*}
   \centering
   \includegraphics[width=6cm]{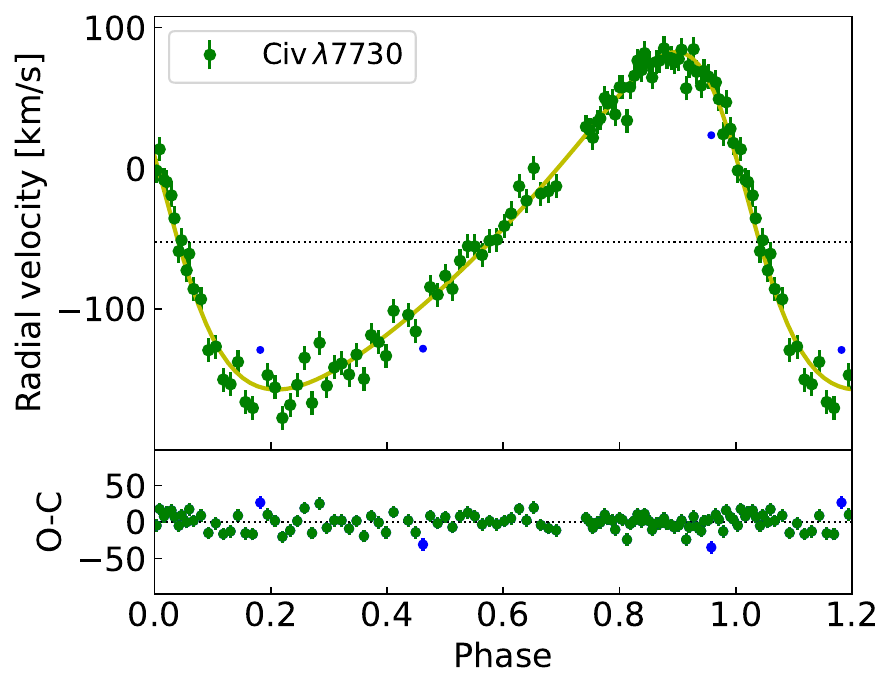}
   \includegraphics[width=6cm]{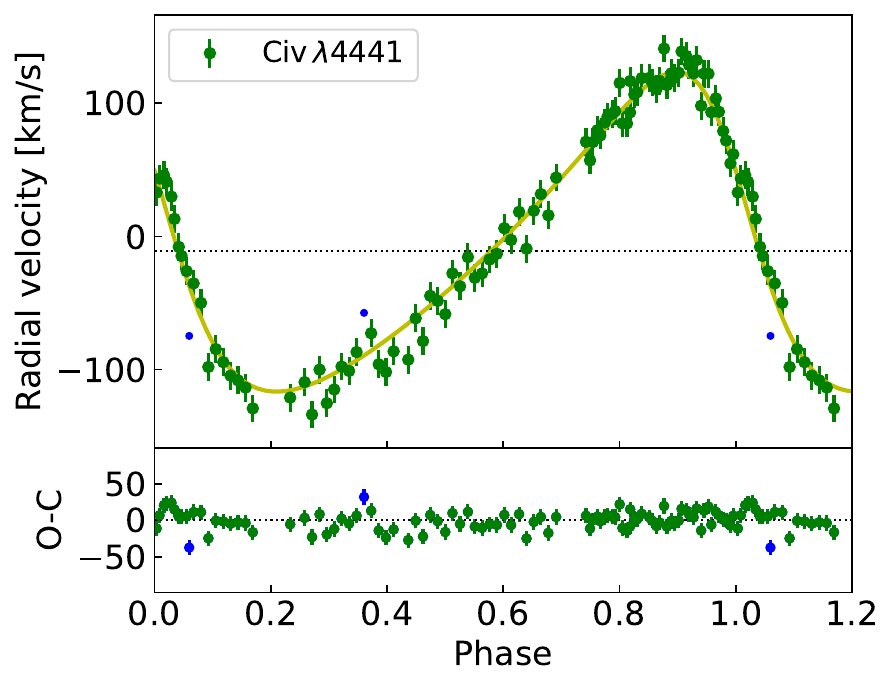}
   \includegraphics[width=6cm]{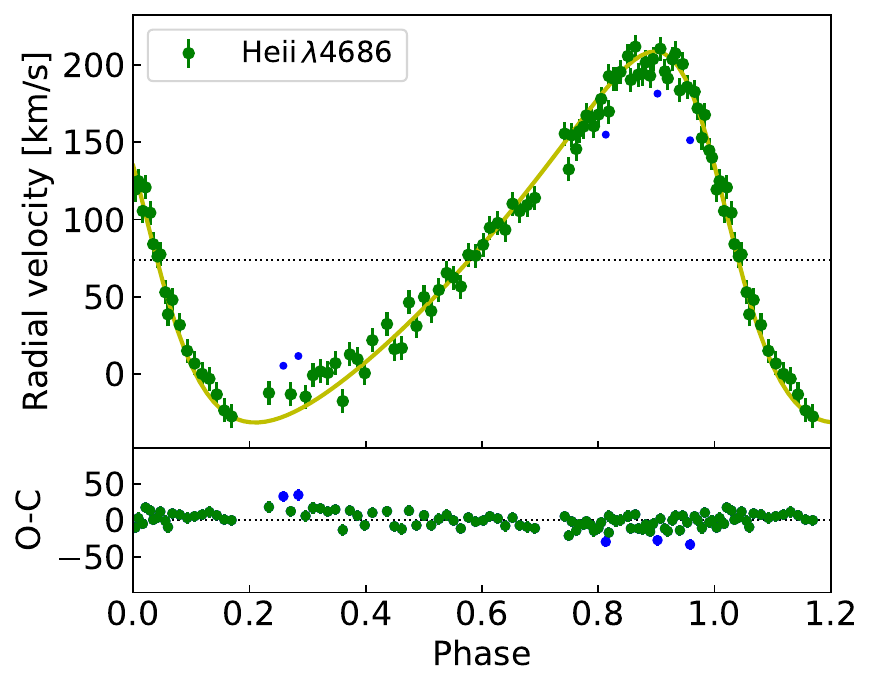}
   \centering
   \caption{As Fig.\ref{Fig:C1} but for three lines that fit the predicted radial velocity solution.
           }
              \label{Fig:C2}%
    \end{figure*}
\begin{figure*}
   \includegraphics[width=6cm]{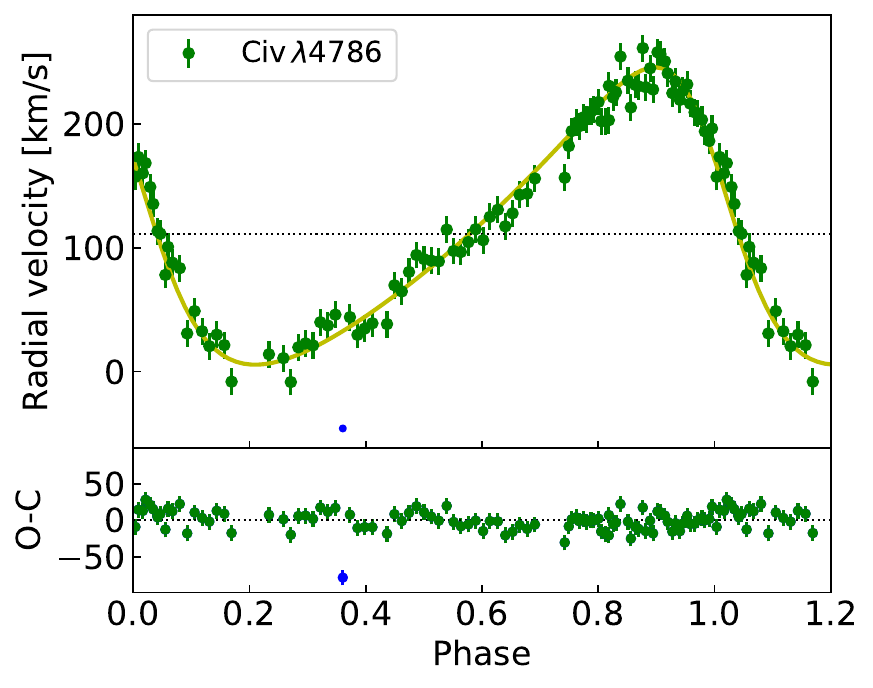}
   \includegraphics[width=6cm]{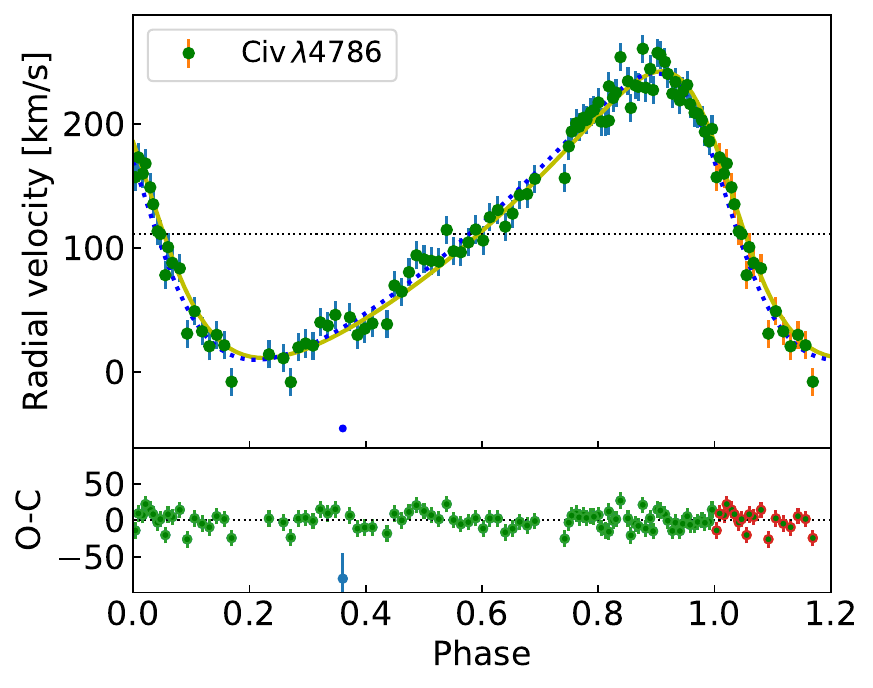}
   \centering
   \caption{Left panel: As Fig.\ref{Fig:C2}  for the forth line which fits the predicted radial velocity solution. Right panel: The solution allowing for an apsidal motion with the parameters given in  
   Tab.\,\ref{Tab:omegadot}. The blue dotted line is the orbital velocity curve without an apsidal motion and identical to the yellow orbit line in the left panel. The difference between the two solutions is very small and hardly visible in the O-C diagram.
           }
              \label{Fig:C3}%
    \end{figure*}

The radial velocity curves of emission lines in WR binaries are well-known to depend on the particular atomic transition being considered.
For a given system, the differences in its RV curves derived from  different lines can be traced to the presence of periodic line profile variations in the measured emission lines, a phenomenon that has been known at least since the 1950s \citet{Munch1950}.
With the orbital parameters relatively precisely determined in sections 3 and 4 we re-evaluate the radial velocity measurements in order to learn which lines are preferably used for an orbit determination. 

\citet{Schmutz_etal1997} have measured the radial velocities of seven emission lines of $\gamma^2$ Vel over more than an orbit in 1996 and they have given the orbital solutions for these lines in their Table\,1.
There is an apparent scatter of the two parameters $e$ and $\omega$, which is due to a strong co-dependency of the two parameters. Only their combination $e \sin(\omega)$ can be obtained reliably from radial velocities.
In table\,\ref{radvel} we have re-solved the orbital solution as given in  Table\,1 of \citet{Schmutz_etal1997} but instead of solving for $e$ and $\omega$ we have solved for $e \sin(\omega)$ and $e \cos(\omega)$. This yields a consistent result for $e \sin(\omega)$.
A co-dependency between the periastron date $T_0$ and $e \cos(\omega)$ remains.

In Figs.\,\ref{Fig:C1}, \ref{Fig:C2}, and \ref{Fig:C3} we compare the measured radial velocities to an orbit predicted by the orbital parameters obtained from the astrometric observations as given in the column with $P=78.524$\,d of Tab.\,\ref{tab:orbit} and adopting a velocity amplitude $K=120$\,km/s. This value results from the estimated sum of amplitudes as derived in Sect.\,\ref{Sec:5}, $K_\mathrm{est}=155.6\pm 0.6$\,km/s, which is $4.8$\,km/s less than the sum of radial velocity amplitudes derived by \citet{Schmutz_etal1997}, $K_\mathrm{WR}=122$\,km/s and $K_\mathrm{O}=38.4$\,km/s. 
The adoption of $K_\mathrm{WR}=120$\,km/s is a likely value intermediate between the full correction to apply to the WR amplitude and no correction.
As can be seen (Fig.\,\ref{Fig:C1}) there are three emission lines out of the seven measured by \citet{Schmutz_etal1997} 
whose RV curves are shifted to earlier phases 
and show larger velocity amplitudes than the estimated $K=120$\,km/s. Four lines (Figs.\,\ref{Fig:C2} and \ref{Fig:C3}) fit the predicted radial velocity orbit very well.

The explanation of the shift is thought to come from the distortion due to a disturbing contribution of the wind-wind collision zone as suspected in many papers and investigated by \citet{Richardson_etal2017}. \citet{Schmutz_etal1997} have discussed the shift of the 
C\,{\sc iii/iv}\,$\lambda4652$ line complex and noted that the change of line width might explain the influence. 
The effects are reminiscent of the trigonometric identity  $A_1\sin(\nu)+A_2\cos(\nu)\equiv \surd (A_1^2+A_2^2)\sin(\nu + \Phi)\, ,$ where $\nu$ corresponds to the true anomaly and the phase shift is $\Phi=\arctan(A_2/A_1)$. 
The line He\,{\sc ii}\,$\lambda$4860 belongs to the shifted lines because the O star contributes  hydrogen to the wind-wind collision zone, i.e.\ there is additional $H\beta$ emission.

To the eye, all four of the fitted lines seem to agree well with the 
astrometric RV curve. However, for a $\chi^2$-statistics they deviate up to 3\,$\sigma$. It turns out that this disagreement can be removed if the orbit curve is slightly shifted to later phases, which can be explained by an apsidal motion
between 1996 and the recent interferometric observations. We have calculated an apsidal motion by fitting three parameters, $K$, $V_0$, and $\dot\omega$ to the four lines. 

The results are listed in Tab.\,\ref{Tab:omegadot}.
The average amplitude of the four lines is $\bar{K}=118.5\pm 2.6$\,km/s, where the uncertainty is the standard deviation of the four values divided by the square root of three. This value is in agreement with the estimated $K$ discussed in Sect.\,\ref{Sec:5}. However, we note that the individual estimates for the uncertainty of the velocity amplitude are of the order of 1.5\,km/s, significantly less than the standard deviation of the four values. This implies that there might be a systematic effect also influencing the lines, which we selected to be less affected by emission from the wind-wind collision zone.

The average apsidal motion is $\dot{\omega}=(1.6\pm 0.5)\,10^{-4}$\, deg/d, with the uncertainty again the standard deviation of the four values divided by the square root of three. The deduced apsidal motion is a 3\,$\sigma$ result and about 35 times as large as the apsidal motion produced by relativistic effects \citep[Eq.\,1][]{Baroch_etal2021}. However, it is not unlikely that there is a noticeable apsidal motion because of the gravitational influence of nearby cluster stars.

\begin{table*}
      \caption[]{Re-determination of the orbital parameters of $\gamma^2$ Vel from the radial velocities of emission lines as measured by \citet{Schmutz_etal1997} from spectra  obtained in 1996 and used for their Table\,1. The orbital solution was calculated with 
      $e \sin(\omega)$ and $e \cos(\omega)$ as free parameters instead of $e$ and $\omega$. The later two parameters given in the table below are calculated from the solution and shown for information. Also for comparison we have inserted the astrometric solution as last column (from the column with $P=78.524$\,d in Tab.\,\ref{tab:orbit}). }
         \label{radvel}
        \begin{tabular}{lrrrrrrrl}
            \hline
            \noalign{\smallskip}
                       & C\,{\sc iii}/C\,{\sc iv} & C\,{\sc iv} & C\,{\sc iv} & C\,{\sc iv} & C\,{\sc iii} & He\,{\sc ii} & He\,{\sc ii} & astrometric\\
             Parameter & $\lambda 4652$ & $\lambda 7730$ & $\lambda 4441$ & $\lambda 4786$ & $\lambda 6740$ & $\lambda 4686$ & $\lambda 4860$ & solution\\
            \hline
            \noalign{\smallskip}
$T_0$ JD 2,400,000.5+ & 50117.5  & 50122.4  & 50121.4  & 50121.7  & 50118.8  & 50120.7  & 50117.1 & $50120.8\pm 0.3$\tablefootmark{a}\\
$K$ [km/s] & 125.6  & 122.7  & 125.4  & 116.2  & 125.3  & 113.2  & 139.0 & -\\
$V_0$ [km/s] & -20.8  & -52.7  & -13.2  & 110.9  & -71.7  & 75.4  & 11.9 & -\\
$e \sin(\omega)$  & 0.317  & 0.305  & 0.292  & 0.308  & 0.311  & 0.303  & 0.347 & $0.298\pm 0.003$\\
$e \cos(\omega)$  & 0.124  & 0.076  & 0.115  & 0.115  & 0.134  & 0.140  & 0.095 & $0.123\pm 0.001$\\
            \noalign{\smallskip}
$\omega$ [deg] & 68.6  & 75.9  & 68.5  & 69.6  & 66.6  & 65.2  & 74.7 & $67.5\pm 0.3$\\
$e$  & 0.341  & 0.314  & 0.314  & 0.328  & 0.338  & 0.334  & 0.360 & $0.322\pm 0.002$\\
            \noalign{\smallskip}
$\sigma_{O-C}$ [km/s]  & 10.7  & 8.5  & 10.4  & 10.9  & 15.5  & 7.8  & 27.6 \\
            \hline
         \end{tabular}
\tablefoot{
\tablefoottext{a}{The date of periastron at the time of the radial velocity measurements is obtained by subtracting $129\times (78.524\pm 0.002)$\,d from the astrometric periastron date MJD $T_0=60250.44\pm 0.07$. An alternative estimate is by including the average apsidal motion as given in Tab.\,\ref{Tab:omegadot}. This yields a periastron date of $T_0=50120.5\pm 0.5$.} 
}
   \end{table*}

\begin{table*}
      \caption[]{As Tab.\,\ref{radvel} but adopting the parameters $P=78.524$\d, $T_0=60250.44$, $e=0.322$, and $\omega=67.5$ for optimizing three parameters $K$, $V_0$, and $\dot{\omega}$. }
         \label{Tab:omegadot}
        \begin{tabular}{lrrrrrl}
            \hline
            \noalign{\smallskip}
                       &  C\,{\sc iv} & C\,{\sc iv} & C\,{\sc iv} &  He\,{\sc ii} &   \\
             Parameter &  $\lambda 7730$ & $\lambda 4441$ & $\lambda 4786$ & $\lambda 4686$ &  average\\
            \hline
            \noalign{\smallskip}
$K$ [km/s] &  $119.8\pm 1.5$ &  $125.1\pm 1.6$  &  $115.9\pm 1.7$ &  $113.1\pm 1.2$ & $118.5\pm 4.5$\\
$V_0$ [km/s] &  -51.8  & -13.1  & 111.2  &  75.6  & \\
$ \dot{\omega}$ [deg/d] &  $4.5\,10^{-5}$  & $2.1\,10^{-4}$   & $2.5\,10^{-4}$  &  $1.3\,10^{-4}$ & $(1.6\pm 0.8)\,10^{-4}$\\
            \noalign{\smallskip}
$\sigma_{O-C}$ [km/s]  &  10.5  & 10.6  & 11.3  &  7.8   \\
            \hline
         \end{tabular}
   \end{table*}

\end{appendix}
\end{document}